\documentclass[preprint2]{aastex}
\usepackage{epsfig}

\begin{document}

\title{Measuring the cosmic evolution of dark energy with baryonic
oscillations in the galaxy power spectrum}

\author{Karl Glazebrook} \affil{Department of Physics \& Astronomy,
  Johns Hopkins University, Baltimore, MD 21218-2686, United States}
\email{kgb@pha.jhu.edu}

\author{Chris Blake} \affil{Department of Physics \& Astronomy,
  University of British Columbia, 6224 Agricultural Road, Vancouver,
  B.C., V6T 1Z1, Canada} \email{cab@astro.ubc.ca}

\begin{abstract}

We use Monte Carlo techniques to simulate the ability of future large
high-redshift galaxy surveys to measure the temporal evolution of the dark
energy equation-of-state $w(z)$, using the baryonic acoustic oscillations in the
clustering power spectrum as a `standard ruler'.  Our analysis only
utilizes the oscillatory component of the power spectrum and not its
overall shape, which is potentially susceptible to broadband tilts induced
by a host of model-dependent systematic effects.  Our results are
therefore robust and conservative.  We show that baryon oscillation constraints
can be thought of, to high accuracy, as a  direct probe of the distance --- redshift and expansion rate --- redshift
relations where distances are measured in units of the sound horizon. Distance precisions
of 1\% are obtainable for a fiducial redshift survey covering $10{,}000$
deg$^2$ and redshift range $0.5 < z < 3.5$. 
If  the dark energy is further
characterized by $w(z) = w_0 + w_1 z$ (with a cut-off in the evolving term
at $z = 2$), we can then measure the parameters
$w_0$ and $w_1$ with a precision exceeding current knowledge by a factor
of ten: $1\sigma$ accuracies $\Delta w_0 \approx 0.03$ and $\Delta w_1
\approx 0.06$ are obtainable (assuming a flat universe and that the other
cosmological parameters $\Omega_{\rm m}$ and $h$ could be measured
independently to a precision of $\pm 0.01$ by combinations of future CMB
and other experiments).  We quantify how this performance degrades with
redshift/areal coverage and knowledge of $\Omega_{\rm m}$ and $h$, and
discuss realistic observational prospects for such large-scale
spectroscopic redshift surveys, with a variety of diverse techniques.  We
also quantify how large photometric redshift imaging surveys could be
utilized to produce measurements of $(w_0,w_1)$ with the baryonic
oscillation method which may be competitive in the short term. 
\end{abstract}

\section{Introduction}

In the current standard cosmological model, baryonic matter and cold dark
matter together contribute only about a third of the total energy density
of the Universe.  One of the most important puzzles in cosmology is to
account for the remaining two-thirds of the energy, which is required to
render the Universe approximately spatially flat, as demanded by recent
observations of the Cosmic Microwave Background \citep[CMB;][]{deBer00}. 

The existing set of cosmological data -- consisting principally of
measurements of the CMB, of the local clustering of galaxies, and of
distant supernovae -- can be understood by invoking the existence of the
`cosmological constant' $\Lambda$ originally envisaged by Einstein, such
that it contributes a present-day fractional energy density
$\Omega_\Lambda = 0.73 \pm 0.04$ \citep{Sper03}.  The remarkable
consequence of this model is that $\Lambda$ acts as a `repulsive gravity',
driving the rate of cosmic expansion into a phase of acceleration. 
Equally surprisingly, this acceleration has been observed reasonably
directly by an apparent dimming of distant supernovae
\citep{Riess98,Perl99}. 

The cosmological constant component is naturally attributed to the
inherent energy density of the vacuum.  However, the `expected'
quantum-mechanical Planck energy density is larger than that required to
account for the accelerating rate of cosmic expansion by an exceptionally
large dimensionless factor of order $c^5\, G^{-1} \hbar^{-1} H_0^{-2}$
$\sim 10^{122}$. This profound difficulty has motivated the development of
alternative models for the `dark energy' -- i.e.\ the causative agent of
accelerating expansion.  Many of these models, such as `quintessence'
\citep{RP88}, feature a {\it dynamic} component of dark energy whose
properties evolve with time (e.g.\ a rolling scalar field $\phi$).  These
predictions are commonly characterized in terms of the dark energy
equation-of-state $w(z) = P/\rho$, relating its pressure $P$ to its energy
density $\rho$ (in units where the speed-of-light $c = 1$).  For the
cosmological constant model, $w(z) = -1$ at all epochs. 

These competing models for the dark energy are essentially untested,
because the current cosmological dataset is not adequate for delineating
variations in the function $w(z)$ with redshift, which is the essential
requirement for distinguishing quintessential cosmologies from a
cosmological constant.  New experiments are demanded, which must
be able to recover the characteristics of dark energy with unprecedented
precision, including any cosmic evolution.  The study of the nature of
dark energy is the current frontier of observational cosmology, and has
been widely identified as having profound importance for physics as a
whole \citep{QuarkCosmos}. 

Given the fundamental importance of accelerating cosmic expansion and the
possibility of confounding systematic effects in the supernova data, other
precision probes of the dark energy model are clearly desirable.  In
\cite{BG03}, hereafter Paper I, we suggested that the small-amplitude
`baryonic oscillations', which should be present in the power spectrum of
the galaxy distribution on large scales ($\gtrsim 30$ Mpc), could be used
as a `standard cosmological ruler' to measure the properties of dark
energy as a function of cosmic epoch, provided that a sufficiently-large
high-redshift ($z \ga 1$) galaxy survey was available \citep[see
also][]{SE03,Linder03,HH03}. The characteristic sinusoidal ruler scale
encoded by the baryonic oscillations is the sound horizon at
recombination, denoted $s$.  This length scale is set by straight-forward
linear physics in the early Universe and its value is determined
principally by the physical matter density ($\rho_m \propto \Omega_m
h^2$).  The cosmological uncertainty in this parameter combination is
ameliorated by an advantageous cancellation of pre-factors of $\Omega_m
h^2$ in the ratio of low-redshift distances to $s$ (see Paper I).  The
residual dependence of $s$ on $\Omega_m$ and other cosmological parameters
is small \citep{EW04}, rendering observations of the acoustic oscillations
a powerful geometric probe of the cosmological model. 

This acoustic signature has recently been identified at low redshift in
the distribution of Luminous Red Galaxies in the Sloan Digital Sky Survey
\citep{Dan05} \citep[see also][]{Cole05}.  Although these data are
insufficient for precise measurements of the dark energy, this analysis
represents a striking validation of the technique.  The challenge now is
to create larger and deeper surveys.  In Paper I we demonstrated that,
given a galaxy redshift survey at $z \sim 1$ mapping a total cosmic volume
several times greater than that of the Sloan main spectroscopic survey in
the local Universe (we define $V_{\rm Sloan} \equiv 2 \times 10^8 \,
h^{-3}$ Mpc$^3$), the equation-of-state of the dark energy could be
recovered to a precision $\Delta w \approx 0.1$ (assuming a model in which
$w$ is constant). The precision of this experiment scales with cosmic
volume $V$ in a predictable manner (roughly in accordance with
$1/\sqrt{V}$) and it is not unfeasible to imagine an ultimate `all-sky'
high-redshift spectroscopic survey within $\sim 20$ years. 

We believe that this standard ruler technique would powerfully complement
proposed future supernova searches such as the SNAP project \citep[e.g.][
{\tt http://snap.lbl.gov}]{Ald02} permitting, for example, direct tests of
the `reciprocity relation' which predicts that the true luminosity
distance $d_L(z)$ (measured by a standard candle) is exactly the same as
the angular diameter distance $d_A(z)$ (measured by a standard ruler), up
to a factor of $(1+z)^2$ \citep{BK04}.  Furthermore, we argue that it is
{\it not yet conclusively proven} that the dimming of supernovae caused by
cosmic acceleration can be distinguished with sufficient accuracy from
other possible systematic effects, such as changes in the
intrinsic properties of supernovae with galactic environment (e.g.\
metallicity), dust extinction effects, population drift and the difficulties of
sub-percent-level photometric calibration (including K-corrections) across
wide wavelength ranges. It is
therefore important to pursue alternative precise high-redshift probes of
the cosmological model.  A particular advantage of the baryonic
oscillations method is that it is probably free of major systematic
errors, assuming that galaxy biasing on large scales is not pathological. 

In Paper I we developed a Monte Carlo, semi-empirical approach to
transforming synthetic galaxy redshift surveys into constraints on the
value of $w$. This contrasts with, and complements, other analytical
approaches to the problem such as those using Fisher-Information
techniques \citep[e.g.][]{SE03,HH03}.  In this present study we extend and
refine our methodology to the more general case where we allow the
equation-of-state of dark energy to have a dependence on redshift, $w(z) =
w_0 + w_1 z$, and we recover joint constraints upon $(w_0,w_1)$. 
Furthermore, we apply the standard ruler independently to the radial and
tangential components of the power spectrum, which has the effect of
producing separate measurements of the co-moving 
angular-diameter distance to the effective
redshift of each survey slice, $x(z)$, and the rate of change of this
quantity with redshift, $x'(z) \equiv dx/dz$.  We note for clarity that in a flat Universe:
$$ x(z) = D_A(z) (1+z) $$
$$x'(z) = c/H(z)$$
where $D_A$ is the physical angular diameter distance and $H(z)$ is the Hubble
factor (Universal expansion rate) at redshift $z$.
We also examine in more
detail the effect of uncertainties in cosmological parameters such as the
matter density and the Hubble parameter.  In addition, we present an
in-depth discussion of the observational requirements and prospects for
realistic surveys, based upon both spectroscopic redshifts and photometric
redshifts. 

The plan of this paper is as follows: in Section~\ref{secmeth} we give a
very detailed description of our methodology, greatly expanding on Paper
I, and the approximations we made to allow us to simulate a range of large
surveys, and in Section~\ref{secdetect} we quantify the size of redshift
survey required to detect the oscillatory component of the power spectrum. 
The recovered constraints on $(w_0, w_1)$ for various simulated surveys
are presented in Section~\ref{secmeas} for spectroscopic surveys and in
Section~\ref{secphoto} for photometric-redshift surveys.  Finally, in an
Appendix we present the results of a very large computation designed to
test the effects of our approximations. 

\section{Monte Carlo Methodology} 
\label{secmeth}

Our methodology for simulating future galaxy redshift surveys and
assessing their efficacy for measuring acoustic oscillations was
summarized in Paper I.  In this Section we provide a more detailed account
of our procedures.  In addition, we have implemented various extensions to
the methodology of Paper I: 

\begin{enumerate}

\item We now fit separate acoustic oscillation scales in the
  tangential and radial directions. In Paper I we fitted to the
  angle-averaged power spectrum, effectively assuming that the shift
  in the apparent radial and transverse scales as the cosmology was
  perturbed about the fiducial value were the same.  This is only
  approximately true for $z\sim 1$ and breaks down at low and high
  redshift (see Figure~5 of Paper~I). In the new approach we make use
  of the different redshift dependencies of radial and transverse
  scales to provide extra cosmological constraints increasing
  signal:noise.  Specifically, the tangential scale is controlled by
  the co-moving distance to the effective redshift of the survey,
  $x(z)$, and the radial scale is determined by the rate of change of
  this quantity with redshift, $x'(z) \equiv dx/dz = c/H(z)$, where
  $H(z)$ is the Hubble constant measured by an observer at redshift
  $z$.  This is useful because $H(z)$ is directly sensitive to the
  dark energy density at the redshift in question.

\item We allow the equation-of-state of dark energy $w(z)$ to have a
  dependence on redshift, $w(z) = w_0 + w_1 z$, and we recover joint
  constraints upon $(w_0,w_1)$.  We do not claim that this expression
  faithfully describes dark energy in the real Universe.  In
  particular, models with $w_1 > 0$ become unphysical at high redshift
  unless we impose a cut-off for the evolving term: we assume that
  $w(z > z_{\rm cut}) = w_0 + w_1 z_{\rm cut}$ where $z_{\rm cut} =
  2$, and we ensure that matter dominates at high redshift, i.e. $w_1
  \le (-w_0)/z_{\rm cut}$.  However, usage of the equation $w(z) = w_0
  + w_1 z$ facilitates comparison with other work such as \cite{SE03}
  who use the same parameterization, empirically describes a range of
  dark energy models \citep{WA02}, and permits a first disproof of the
  cosmological constant scenario, if $w_1 \ne 0$ and/or $w_0 \ne -1$.
  We note that the alternative parameterization $w(z) = w_0 + w_a
  (1-a)$, where $a=(1+z)^{-1}$ is the usual cosmological scale factor,
  encodes a more physically reasonable behaviour at high redshift
  \citep{Linder02}.  In this paper we compute $(w_0,w_a)$ constraints
  for one case.

\item We place Gaussian priors upon the other relevant cosmological
  parameters, the matter density $\Omega_{\rm m}$ and the Hubble
  parameter $h = H(z=0)/(100$ km s$^{-1}$ Mpc$^{-1})$, rather than
  assuming that their values are known precisely.

\end{enumerate}

\subsection{Method Summary}
\label{secsumm}

As described in Paper I, the philosophy of our analysis is to maintain
maximum independence from models.  When measuring the acoustic
oscillations from simulated data, we divide out the overall shape of the
power spectrum via a smooth `reference spectrum'.  We then fit a simple
empirically-motivated decaying sinusoid to the remaining oscillatory
signal.  Hence we do not utilize any information encoded by the shape of
the power spectrum.  This shape may be subject to smooth broad-band
systematic tilts induced by such effects as complex galaxy biasing
schemes, quasi-linear growth of structure, a running primordial spectral
index, and redshift-space distortions.  However, it would be very
surprising if any of these phenomena introduced {\it
  oscillatory} features in $k$-space liable to obscure the distinctive
acoustic peaks and troughs.  We note that any model where the
probability of a galaxy forming depends only on the local density
field leads to linear bias on large scales \citep{Coles93,SW98}.
Furthermore, linear biassing is observed to be a very good
approximation on large scales \citep[e.g.][]{PeaDod,Cole05}.
This is in agreement with numerical
simulations of galaxy formation which show that galaxies and/or massive haloes
faithfully reproduce the
acoustic oscillations \citep{Spr05,Ang05}.

Of course, a full power spectrum template should be fitted to real data as
well: our aim here is to derive robust, conservative lower limits to the
efficacy of baryon oscillations experiments, using only the information
contained in the oscillations. 

An important point is that the fractional error in the measured galaxy
power spectrum, $\sigma_P/P$, after division by a smooth overall fit, is
independent of the absolute value of $P(k)$ if the error budget is
dominated by cosmic variance rather than by shot noise.  In this sense, an
incorrect choice of the underlying model power spectrum in our simulations
does not seriously affect the results presented here. Having secured a
detection of the acoustic signature, if one is then prepared to model the
underlying power spectrum -- correcting for such systematic effects as
non-linear gravitational collapse, redshift-space distortions and halo
bias -- then more accurate constraints on cosmological parameters would
follow \citep[see][]{Dan05}. 

In summary (see Section \ref{secsteps} for a more detailed account): we
generate a model matter power spectrum in the linear regime using the
fitting formulae of \cite{EH98}, assuming a primordial spectral index $n =
1$ (as suggested by inflationary models) and fiducial cosmological
parameters $\Omega_{\rm m} = 0.3$, $h = 0.7$ and baryon fraction
$\Omega_{\rm b}/\Omega_{\rm m} = 0.15$, broadly consistent with the latest
determinations \citep[e.g.][]{Sper03}.  In Paper~I we showed that the
cosmological constraints are fairly insensitive to the exact value of
$\Omega_{\rm b}/\Omega_{\rm m}$ (Figures~7--8 in Paper~I; increasing
$\Omega_{\rm b}$ results in a somewhat higher baryonic oscillation
amplitude, hence a more precise measure of the standard ruler).  We assume
that the shape of $P(k)$ does not depend on the dark energy component, and
take the $z = 0$ normalization $\sigma_8 = 1$.  The model $P(k)$ is then
used to generate Monte Carlo realizations of a galaxy survey covering a
given geometry, deriving redshifts and angular co-ordinates for the
galaxies using a fiducial flat cosmological constant model.  The
realizations are then analyzed for a grid of assumed dark energy models. 
$P(k)$ is measured using a Fast Fourier Transform (FFT) up to a maximum
value of $k$ determined by a conservative estimate of the extent of the
linear regime at the redshift in question (see Paper I, Figure 1).  The
measured power spectrum is fitted with a decaying sinusoid with the
`wavelength' as a free parameter.  By comparing the fiducial wavescale
(determined using the values of $\Omega_{\rm m} h^2$ and $\Omega_{\rm b}
h^2$ in conjunction with a standard fitting formula, e.g. \cite{EB99})
with the distribution of fitted wavescales across the Monte Carlo
realizations, we can reject each assumed dark energy model with a
derivable level of significance.  We assume a flat universe (in Section~\ref{secmethlik}, Figure
\ref{figcurvgen} below we indicate how our dark energy measurements weaken with
declining knowledge of $\Omega_{\rm k}$). 

\begin{figure}[htbp]
\begin{center}
\epsfig{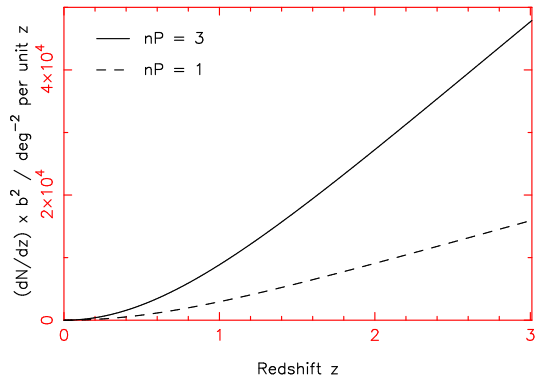}
\end{center}
\caption{The surface density of galaxies per unit redshift, $dN/dz$,
  required to achieve power spectrum shot noise levels $n \times P =
  1$ (dashed line; equal error contribution due to shot noise and to
  cosmic variance) and $n \times P = 3$ (solid line; adopted in our
  simulations).  The power spectrum $P$ is evaluated at a `typical'
  scale $k = 0.2 \, h$ Mpc$^{-1}$; at zero redshift this value is
  taken to be $P = 2500 \, h^{-3}$ Mpc$^3$ (and is scaled to higher
  redshifts using the square of the linear growth factor $D_1(z)$).  If
  the galaxies possess a linear bias $b$, then the amplitude of their
  power spectrum scales as $b^2$ and a proportionately lower surface
  density is required.  For $b=1$ galaxies in a redshift slice of
  width $\Delta z = 0.2$ at $z=1$, a density $\approx 1800$ deg$^{-2}$
  is required to achieve $n \times P = 3$.}
\label{fignp}
\end{figure}

Throughout this paper we ensure that our model galaxy surveys contain
sufficient objects that the contribution of shot noise to the error in the
power spectrum is sub-dominant to that of cosmic variance.  In an
analytical treatment \citep[e.g.][]{Teggers97}, the relative contributions
of shot noise and cosmic variance can be conveniently expressed in terms
of the quantity $n \times P$, where $n$ is the typical number density of
galaxies in the survey volume and $P$ is the galaxy power spectrum
evaluated at some typical scale measured by the survey.  Analytically, the
errors due to shot noise and to cosmic variance are equal when $n \times P
= 1$.  For the simulations described in this paper, we uniformly populated
the survey volume with sufficient galaxies that $n \times P = 3$, where
$P$ is evaluated at a characteristic scale $k = 0.2 \, h$ Mpc$^{-1}$.  The
surface density of galaxies required to achieve this condition is
illustrated in Figure \ref{fignp}. 

\subsection{Detailed fitting methodology}
\label{secsteps}

Here we outline the Monte-Carlo approach we have implemented in our code. 
In fact for most of our analysis we utilized a simple approximation as
explained in Section \ref{secstream}: to be specific we omitted steps 8
and 9 below, which are very expensive in computational resources (we test
this approximation in the Appendix). 

\begin{enumerate}

\item A fiducial cosmology is chosen for the simulation: for example
  $(\Omega_{\rm m},h,w_0,w_1) = (0.3,0.7,-1,0)$.  A fiducial baryon
  fraction is selected ($\Omega_{\rm b}/\Omega_{\rm m} = 0.15$).

\item A survey redshift range $(z_{\rm min},z_{\rm max})$ and solid
  angle is specified: for example $1.0 < z < 1.3$ and $1000$ deg$^2$.
  The survey sky geometry is assumed to be bounded by lines of
  constant right ascension and declination of equal angular lengths.
  The three-dimensional geometry is therefore `conical' and the
  convolving effect of this window function is included.

\item Using the fiducial cosmology, a cuboid for FFTs is constructed whose
  sides $(L_x, L_y, L_z)$ are just sufficient to bound the survey
  volume. Note, only the enclosed cone is  populated by galaxies in order that
  the window function effect is treated properly.

\item A model matter power spectrum $P_{\rm mass}(k,z=0)$ is computed
  for the chosen parameters $(\Omega_{\rm m},\Omega_{\rm b},h)$ from
  the fitting formula \citep{EH98}, assuming a $z=0$ normalization
  $\sigma_8 = 1$ and a primordial power-law slope $n=1$.  The survey
  slice is assumed to have an `effective' redshift $z_{\rm eff} =
  (z_{\rm min} + z_{\rm max})/2$.  The power spectrum is scaled to
  this redshift using the linear growth factor $D_1(z)$, obtained by
  solving the full second-order differential equation \citep[see
    e.g.][]{LinJen} to enable us to treat non-$\Lambda$CDM
  cosmologies:
\begin{equation}
P_{\rm gal}(k,z_{\rm eff}) = P_{\rm mass}(k,0) \, D_1(z_{\rm eff})^2
\, b^2
\end{equation}
where we use a constant linear bias factor $b$ for the clustering of
galaxies with respect to matter.  The value $b=1$ is assumed for our
surveys, unless otherwise stated.

\item A set of Monte Carlo realizations (numbering 400 for all
  simulations presented here) is then performed to generate many
  different galaxy distributions consistent with $P_{\rm gal}(k)$, as
  described in steps 6 and 7.

\item A cuboid of Fourier coefficients is constructed with grid lines
  set by $dk_i = 2\pi/L_i$, with a Gaussian distribution of amplitudes
  determined from $P_{\rm gal}(k)$, and with randomized phases.  The
  gridding is sufficiently fine that the Nyquist frequencies in all
  directions are significantly greater than the smallest scale for
  which a power spectrum is extracted (i.e.\ the linear/non-linear
  transition scale at the redshift $z_{\rm eff}$).

\item The Fourier cuboid is FFTed to determine the density field in
  the real-space box.  This density field is then Poisson sampled
  within the survey `cone' to determine the number of galaxies in each
  grid cell.

\item Using the fiducial cosmological parameters, this distribution is
  converted into a simulated catalogue of galaxies with redshifts and
  angular positions, for each Monte Carlo realization.

\item We now assume a trial cosmology: for example $(\Omega_{\rm
  m},h,w_0,w_1) = (0.3,0.7,-0.9,0)$.  The co-moving co-ordinates of
  the galaxies are computed in the trial cosmology as would be done by
  an observer without knowledge of the true cosmology.

\item The power spectrum of the simulated survey for the trial
  cosmology is measured using standard estimation tools \citep{FKP94}.
  Power spectrum modes in Fourier space are divided into bins of
  $(k_{\rm perp},k_{\rm par})$ where, if the $x$-axis is the radial
  direction, $k_{\rm par} = k_x$ and $k_{\rm perp}^2 = k_y^2 + k_z^2$.

\item An error bar is assigned to each power spectrum bin using the
  variance measured over the Monte Carlo realizations.  Note that the
  distribution of realizations also encodes any covariances between
  different power spectrum bins, although the scale of correlations in
  $k$-space is expected to be very small (compared to the separation
  of the acoustic peaks) for the very large survey volumes considered
  here.

\item The measured $P(k_{\rm perp},k_{\rm par})$ is divided by a
  smooth `reference spectrum' following Paper I \citep[i.e. the
    `no-wiggles' spectrum of][]{EH98}, and the result is fitted with a
  simple empirical formula, modified from Paper I to permit separate
  fitting of the sinusoidal scale in the radial and tangential
  directions:
\begin{eqnarray}
&\,& \frac{P(k_{\rm perp},k_{\rm par})}{P_{\rm ref}} = 1 + \nonumber
\\ &\,& \hspace{-1cm} A \, k \, \exp{ \left[ - \left( \frac{k}{0.1 \,
h \, {\rm Mpc}^{-1}} \right)^{1.4} \right] } \times \nonumber \\ &\,&
\hspace{-1cm} \sin{ \left( 2 \pi \sqrt{ \left( \frac{k_{\rm
perp}}{\lambda_{\rm perp}} \right)^2 + \left( \frac{k_{\rm
par}}{\lambda_{\rm par}} \right)^2 } \right) }
\end{eqnarray}
where $k^2 = k_{\rm perp}^2 + k_{\rm par}^2$.  The free parameters are
then the tangential and radial sinusoidal wavescales $(\lambda_{\rm
  perp},\lambda_{\rm par})$ together with the overall amplitude $A$.

\end{enumerate}

\begin{figure}[htbp]
\begin{center}
\epsfig{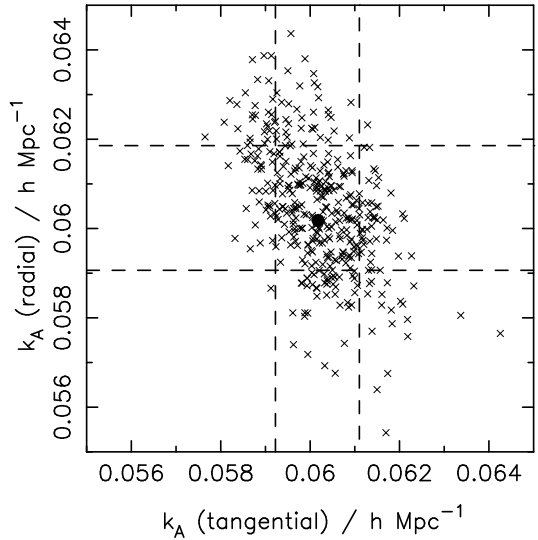}
\end{center}
\caption{Distribution of fitted tangential and radial wavescales for
  400 Monte Carlo realizations of a survey covering volume $10 \,
  V_{\rm Sloan}$ and redshift range $0.75 < z < 1.25$.  The solid
  circle marks the expected value of the acoustic scale and the dashed
  lines bound regions containing $68\%$ of the fitted values.  The
  tangential wavescale may be determined more accurately than the
  radial wavescale because more tangential modes are available
  (i.e.\ a cylindrical shell of radius $k_{\rm perp} = \sqrt{k_y^2 +
    k_z^2}$ centred upon the radial ($x$-) axis).}
\label{figlamfit}
\end{figure}

We can now assign a probability to the trial cosmology.  The Monte Carlo
realizations produce a distribution of fitted wavescales $\lambda_{\rm
perp}$ and $\lambda_{\rm par}$ (Figure \ref{figlamfit}) for the trial
cosmology.  Using these trial cosmological parameters, we can determine
the length of the characteristic ruler $\lambda_{\rm theory}$ using a
standard fitting formula for the sound horizon integral (Equation 1, Paper
1) in terms of $\Omega_{\rm m}$, $\Omega_{\rm b}$ and $h$ \citep{EB99}.
(We remind the reader that  $\lambda_{\rm theory}$ is set in the early Universe
and is insensitive to dark energy parameters.)
The location of the value of $\lambda_{\rm theory}$ in the distribution of
tangential (radial) wavescales over the Monte Carlo realizations allows us
to assign a probability for the trial cosmological parameters.  For
example: if $\lambda_{\rm theory}$ lies at the $16^{\rm th}$ percentile of
the distribution, the rejection probability is $2 \times (50-16) = 68\%$.
Note that the simulated observer does not need to know the fiducial
cosmological parameters (including the dark energy model) to perform this
analysis with real data. 

\subsection{The streamlined approach}
\label{secstream}

The full methodology outlined above is too computationally intensive for
the exploration of a full grid of trial cosmological parameters
$(\Omega_{\rm m},h,w_0,w_1)$.  In practice we pursue a streamlined
approach that adopts some simple approximations.  In the Appendix we use a
test case to demonstrate that the results are equivalent to the
utilization of the full methodology. 

In our streamlined approach, we exploit the fact that the accuracy of
measurement of $\lambda_{\rm perp}$ is a very good approximation of the
precision with which we can recover the quantity $x(z_{\rm eff})/s$, where
$x(z_{\rm eff})$ is the co-moving radial distance to the effective
redshift of the survey and $s$ is the value of the sound horizon at
recombination.  This is because (in the flat-sky approximation) the value
of $x$ controls physical tangential scales in the slice (as displacements
$\Delta r$ are governed by $\Delta r = x \times \Delta \theta$). 
Similarly, the accuracy of measurement of $\lambda_{\rm par}$ is
equivalent to that of $x'(z_{\rm eff})/s$, where $x'(z) \equiv dx/dz =
c/H(z)$ (since $\Delta r = x' \times \Delta z$).  The value of $s$ appears
in the denominators because a systematic shift in the standard ruler scale
implies a similar variation in the recovered physical scales $x$ and
$dx/dz$: cosmic distances are measured in units of the sound horizon at
recombination (equivalently, we may think of this measuring rod as the
distance to the CMB: $s = D_{\rm CMB} \times \theta_A$, where $\theta_A$
is the angular scale separating the CMB acoustic peaks). 

Therefore, rather than re-constructing the galaxy distribution using a
trial cosmology, we instead fitted wavelengths $\lambda_{\rm perp}$ and
$\lambda_{\rm par}$ directly in the fiducial cosmology (i.e.\ omitting
steps 8 and 9 above).  The $68\%$ scatter in these fits across the Monte
Carlo realizations was assigned as a $1\sigma$ Gaussian error in the
values of $x(z_{\rm eff})/s$ and $x'(z_{\rm eff})/s$, respectively.  The
likelihood contours for the trial cosmological parameters were then
deduced using standard expressions for $dx/dz$ and $x$ in terms of
$(\Omega_{\rm m},h,w_0,w_1)$ and a fitting formula for $s$ in terms of
$(\Omega_{\rm m},\Omega_{\rm b},h)$ (see Section \ref{secmethlik}). 

This streamlined approach assumes that:

\begin{enumerate}

\item The scatter in fitted wavelengths is independent of the values
  of the cosmological parameters.  In detail, changing the
  cosmological parameters will alter the cosmic volume surveyed
  between $z_{\rm min}$ and $z_{\rm max}$, and therefore the errors in
  the recovered power spectrum in any bin (and hence the accuracy with
  which the sinusoidal scale may be determined).  However, these
  variations can be neglected for small perturbations about a fiducial
  cosmology.

\item The values of $[x(z_{\rm eff}),x'(z_{\rm eff})]$ control
  tangential and radial scales, respectively.  This statement is exact
  for a flat sky, and holds approximately for the conical geometry
  assumed here (if the survey solid angle is not too large).

\end{enumerate}

These approximations are tested in the Appendix.

\subsection{Likelihoods for dark energy models}
\label{secmethlik}

\begin{figure*}[htbp]
\begin{center}
\epsfig{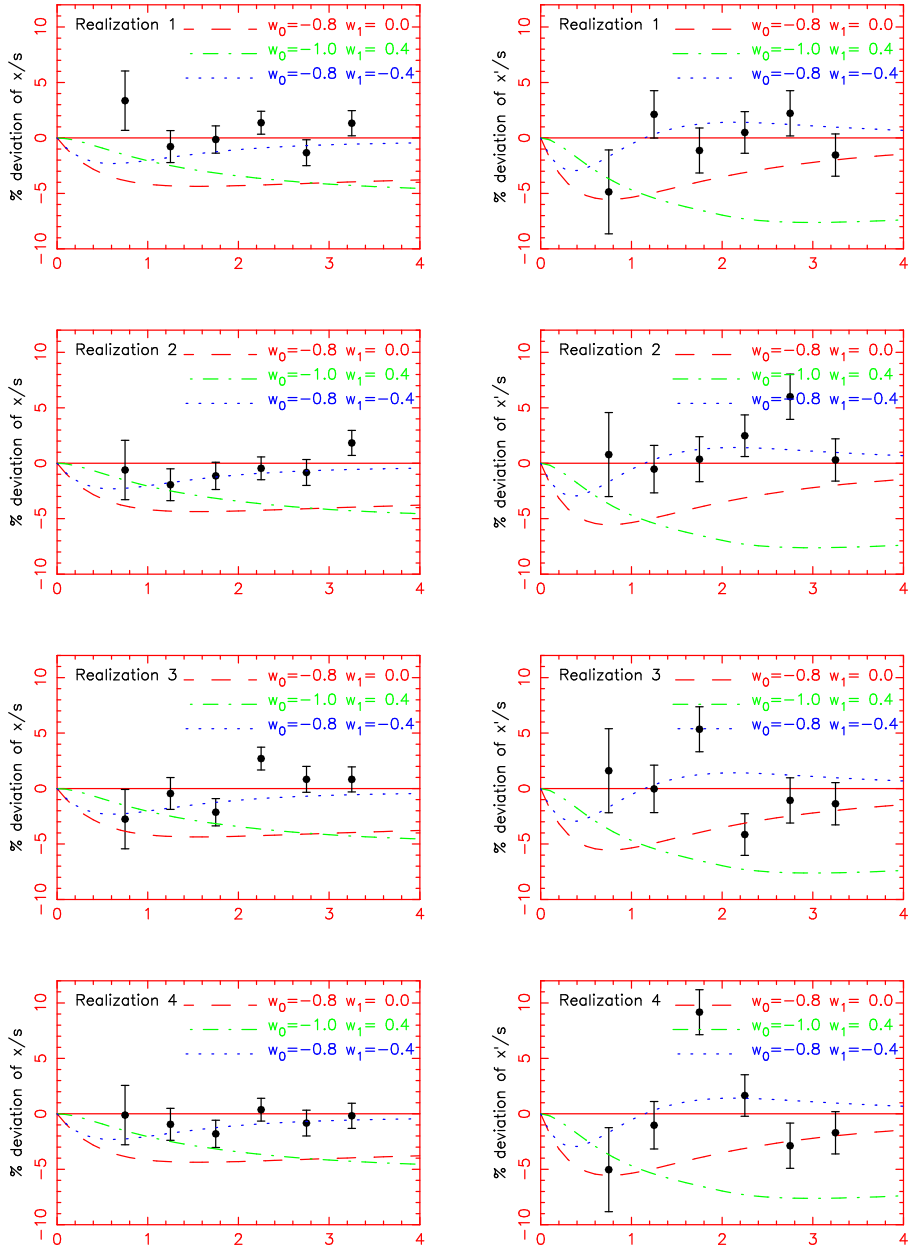}
\end{center}
\caption{Simulated measurements of $x(z)/s$ and $x'(z)/s$ in six
  redshift slices for the first four Monte Carlo realizations of a
  $1000$ deg$^2$ survey.  The plotted values in each redshift bin are
  inferred from the fractional deviation of the fitted tangential and
  radial wavelengths from their fiducial values and the error bars
  represent the standard deviation across the realizations.  The
  various model curves represent the fractional deviations of $x(z)$
  and $x'(z)$ as the dark energy model is varied from the fiducial
  point $(w_0,w_1) = (-1,0)$.  The plots illustrate that these data
  would be sufficient to rule out dark energy parameters $(-0.8,0)$
  and $(-1,0.4)$ with high confidence; but, owing to the cosmic
  degeneracy between $w_0$ and $w_1$, a model $(w_0,w_1) =
  (-0.8,-0.4)$ cannot be confidently excluded.  Note that the
  evolution of $w(z) = w_0 + w_1 z$ is cut off at $z = z_{\rm cut} =
  2$.}
\label{figxdx}
\end{figure*}

The procedure thus far has permitted us to recover values and statistical
errors for the quantities $x(z_{\rm eff})/s$ and $x'(z_{\rm eff})/s$ for
each survey redshift bin.  These measurements are statistically
independent to a good approximation (this is evidenced by the distribution
of fitted wavelengths in Figure \ref{figlamfit} being close to an ellipse
aligned parallel with the axes, displaying only a weak tilt).  Figure
\ref{figxdx} illustrates the simulated recovery of $x(z)/s$ and $x'(z)/s$
for some Monte Carlo realizations of a $1000$ deg$^2$ survey, in redshift
slices of width $\Delta z = 0.5$ from $z = 0.5$ to $z = 3.5$.  Results for
the accuracy of recovery of $x(z)/s$ and $x'(z)/s$ in these redshift
slices are listed in Table \ref{tabxdx}. 

\begin{table*}[htbp]
\begin{center}
\begin{tabular}{ccccc}
\hline
Survey & Area & $z$-bin & Accuracy & Accuracy \\
& (deg$^2$) & & $x/s$ ($\%$) & $x'/s$ ($\%$) \\
\hline
spec-$z$ & 1000 & $0.5 - 1.0$ & 2.7 & 3.8 \\
& & $1.0 - 1.5$ & 1.4 & 2.1 \\
& & $1.5 - 2.0$ & 1.2 & 2.0 \\
& & $2.0 - 2.5$ & 1.0 & 1.9 \\
& & $2.5 - 3.0$ & 1.2 & 2.0 \\
& & $3.0 - 3.5$ & 1.1 & 1.9 \\
\hline
spec-$z$ & 10000 & $0.5 - 0.7$ & 1.7 & 2.7 \\
& & $0.7 - 0.9$ & 1.1 & 2.0 \\
& & $0.9 - 1.1$ & 1.0 & 1.5 \\
& & $1.1 - 1.3$ & 0.7 & 1.4 \\
& & $1.3 - 1.5$ & 0.6 & 1.2 \\
\hline
KAOS & 1000 & $0.5 - 1.3$ & 1.6 & 2.6 \\
& 400 & $2.5 - 3.5$ & 1.2 & 2.3 \\
\hline
photo-$z$ $\sigma_0 = 0.03$ & 2000 & $0.5 - 1.5$ & 2.3 & -- \\
& & $1.5 - 2.5$ & 1.4 & -- \\
& & $2.5 - 3.5$ & 1.3 & -- \\
\hline
\end{tabular}
\end{center}
\caption{Simulated precision of recovery of the quantities $x(z_{\rm
    eff})/s$ and $x'(z_{\rm eff})/s$ from a series of future
  spectroscopic and photometric redshift surveys.  For spectroscopic
  redshift surveys, these quoted accuracies (for area $A_1$) may be
  approximately scaled to other survey areas ($A_2$) by multiplying by
  a factor $\sqrt{A_1/A_2}$ (since the errors in the power spectrum
  measurement $\delta P \propto 1/\sqrt{V}$).  In order to scale the
  photometric redshift results to surveys with different
  r.m.s.\ redshift scatters $\sigma_2$, we can multiply by a further
  factor $\sqrt{\sigma_2/\sigma_1}$ (since the number of usable
  Fourier modes $m$ scales as $1/\sigma$, and $\delta P \propto
  1/\sqrt{m}$).  These simple scalings will break down (1) if the
  redshift range changes significantly, owing to the changing position
  of the non-linear transition, and (2) in the regime where we are
  just resolving the oscillations, when improvement is better than
  $\sqrt{V}$.  The measurement precision of $x/s$ and $x'/s$ is
  determined by Monte Carlo realizations and is accurate to $\pm$ 0.1\%.}
\label{tabxdx}
\end{table*}

We parameterize the dark energy model using an equation-of-state $w(z) =
w_0 + w_1 z$.  The accuracies of $x(z)/s$ and $x'(z)/s$ are then used to
infer joint constraints over a grid of $(w_0,w_1)$, using the standard
formulae for $x'(z) = c/H(z)$ and $x(z) = \int_0^z x'(z') \, dz'$, where
the Hubble constant $H(z)$ at redshift $z$ is a function of $(\Omega_{\rm
m},h,w_0,w_1)$.  For the sound horizon $s$ we used the formulae of
Efstathiou \& Bond (1999, equations 18-20) in terms of $(\Omega_{\rm
m},\Omega_{\rm b},h)$ so we effectively assume that the effect of dark
energy at early times is insignificant.  We explore a range of
uncertainties for these quantities below. For each grid point in the
$(w_0,w_1)$ plane we derive a likelihood for each redshift slice by
marginalizing over Gaussian priors for $\Omega_{\rm m} $ and $\Omega_{\rm
m} h^2$ (the natural variables -- see below) centered upon $\Omega_{\rm m}
= 0.3$ and $h = 0.7$.  The sound horizon is only weakly dependent on
$\Omega_{\rm b}$, therefore we simply fixed the value $\Omega_{\rm b} h^2
= 0.022$ (we checked that the likelihood contours remained unchanged for
reasonable variations in $\Omega_{\rm b} h^2$).  The overall $(68\%,95\%)$
likelihood contours were then determined by multiplying together the
individual likelihoods inferred from the measurements of $x(z)/s$ and
$x'(z)/s$ for each redshift slice. 

\begin{figure}[htbp]
\begin{center}
\epsfig{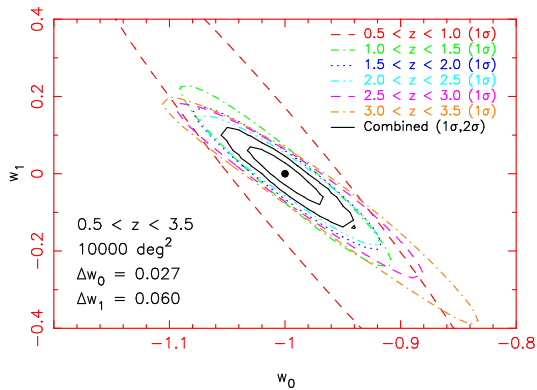}
\end{center}
\caption{Likelihood contours for dark energy model parameters
  $(w_0,w_1)$ for a $10{,}000$ deg$^2$ survey.  Contours are shown for
  six individual redshift slices ranging from $z=0.5$ to $z=3.5$
  ($68\%$), together with the combined result ($68\%$,$95\%$).  For
  this plot, precise knowledge of $\Omega_{\rm m}$ ($= 0.3$), $h$ ($=
  0.7$) and $\Omega_{\rm k}$ ($= 0$) is assumed.  In Figures
  \ref{figpriorsgen} and \ref{figcurvgen} we plot sets of contours for
  different priors in $\Omega_{\rm m}$, $h$ and $\Omega_{\rm k}$.  The
  quoted errors $\Delta w_0$ and $\Delta w_1$ are produced by
  marginalizing over the other dark energy parameter (i.e.\ $w_1$ and
  $w_0$, respectively).  This plot is obtained by scaling the inferred
  measurement accuracies of $x/s$ and $x'/s$ from a $1000$ deg$^2$
  simulation by a factor $\sqrt{10}$.}
\label{figw0w1gen}
\end{figure}

Figure \ref{figw0w1gen} displays the resulting $(w_0,w_1)$ contours for a
$10{,}000$ deg$^2$ survey.  A further approximation has been used to
generate this plot (also tested in the Appendix): that the accuracies of
the fitted wavelengths determined for the $1000$ deg$^2$ simulation
(Figure \ref{figxdx}) may be scaled by a factor $\sqrt{10}$.  This is
simply equivalent to sub-dividing the $10{,}000$ deg$^2$ survey into $10$
separate independently-analyzed pieces.  If we do not make this additional
approximation, unfeasibly large Fourier transforms are required to handle
the size of the survey cuboid. Furthermore, the statistical independence
of $x(z_{\rm eff})$ and $x'(z_{\rm eff})$ in a given redshift slice
becomes weaker, as this independence rests upon the flat-sky
approximation. 

We note that:

\begin{enumerate}

\item There is a significant degeneracy in each redshift slice between
  $w_0$ and $w_1$, because approximately the same cosmology is
  produced if $w_0$ becomes more negative and $w_1$ becomes more
  positive.  The axis of degeneracy is a slow function of redshift,
  which improves this situation somewhat as we combine different
  redshift slices.

\item As redshift increases, the radial oscillations provide
  decreasingly powerful constraints upon the dark energy model,
  because $H(z)$ becomes independent of $(w_0,w_1)$.  This conclusion
  is valid if we are perturbing about the cosmological constant model
  $(-1,0)$, but will not be true in general for models with $w_1 \ne
  0$ for which dark energy may affect dynamics at higher redshift.

\item The tightness of the likelihood contours in the $(w_0,w_1)$
  plane depends significantly upon the fiducial dark energy model (see
  Figure \ref{figfid}).  As $w_0$ and $w_1$ become more positive, dark
  energy grows more influential at higher redshifts and the simulated
  surveys constrain the dark energy parameters more accurately,
  despite the fact that the surveyed cosmic volume is decreasing.

\end{enumerate}

When generating Figure \ref{figw0w1gen}, we assume that the values of
$\Omega_{\rm m}$ ($=0.3$) and $h$ ($=0.7$) are known perfectly. Although
there is some useful cancellation between the trends of the distance scale
$x(z)$ and the standard ruler scale $s$ with the value of $\Omega_{\rm m}
h^2$, there is nevertheless some residual dependence of the experimental
performance on the accuracy of our knowledge of both $\Omega_{\rm m}$ and
$h$, independently. 

In this paper we choose not to combine our results with cosmological
priors from specific proposed experiments (as could be achieved by
combining Fisher matrix information, for example).  We prefer to keep our
presentation in general terms by marginalizing over different priors for
$\Omega_{\rm m}$ and $h$.  We recognize that this approach will not
capture all the parameter degeneracies inherent in future CMB, large-scale
structure or supernova surveys, but nevertheless we can robustly quantify
the accuracy of knowledge required for these other cosmological parameters
such that their uncertainities do not dominate the resulting error in dark
energy parameters.  Combinations with any specific future experiment can
be achieved by using our results listed in Table \ref{tabxdx} which represent
the fundamental observables recovered by this method: $x(z)$ and $x'(z)$ in
units of the sound horizon.

In general terms, one degree of freedom in other parameters is constrained
by the excellent measurement of the physical matter density $\Omega_{\rm
m} h^2$ afforded by the CMB angular power spectrum: accuracies of about
$3\%$ ($\sigma(\Omega_{\rm m} h^2) \approx 0.004$) and $1\%$
($\sigma(\Omega_{\rm m} h^2) \approx 0.001$) are possible with the WMAP
and Planck satellites, respectively \citep{Balbi03}.  However, a second
independent constraint on a combination of $\Omega_{\rm m}$ and $h$ is
also required. 

\begin{figure}[htbp]
\begin{center}
\epsfig{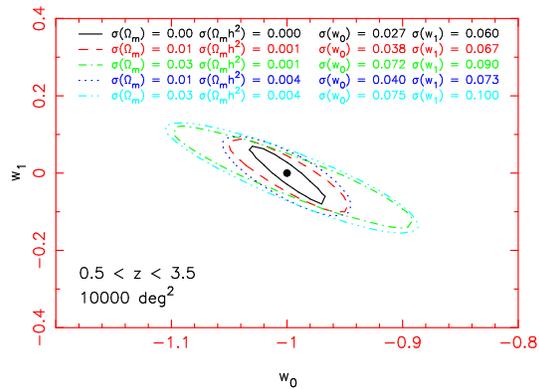}
\end{center}
\caption{Likelihood contours ($68\%$) for dark energy model parameters
  $(w_0,w_1)$ for the same (combined) survey as Figure
  \ref{figw0w1gen}, considering various different Gaussian priors upon
  the values of $\Omega_{\rm m} h^2$ and $\Omega_{\rm m}$, with
  standard deviations $\sigma(\Omega_{\rm m} h^2)$ and
  $\sigma(\Omega_{\rm m})$.  We assume $\Omega_{\rm k} = 0$.  For the
  values of $\sigma(\Omega_{\rm m} h^2)$ we make assumptions
  representative of the WMAP and Planck CMB experiments, for each of
  these cases we further consider $\sigma(\Omega_{\rm m}) = 0.01$ and
  $0.03$.  For a survey of $10{,}000$ deg$^2$ covering redshift range
  $0.5 < z < 3.5$, WMAP suffices to determine the value of
  $\Omega_{\rm m} h^2$, but we require an independent constraint
  $\sigma(\Omega_{\rm m}) \simeq 0.01$ in order that this uncertainty
  is not limiting.}
\label{figpriorsgen}
\end{figure}

This is illustrated by Figure \ref{figpriorsgen}, in which we re-compute
the overall likelihoods in the $(w_0,w_1)$ plane, marginalizing over the
WMAP and Planck errors in $\Omega_{\rm m} h^2$ together with a second
independent Gaussian prior on $\Omega_{\rm m}$. We conclude that for a
survey of $10{,}000$ deg$^2$, $\Omega_{\rm m}$ must be known with an
accuracy $\sigma(\Omega_{\rm m}) \simeq 0.01$ (in conjunction with the
WMAP or Planck determination of $\Omega_{\rm m} h^2$) in order that this
uncertainty is not limiting.  Note that as the cosmic volume surveyed
increases, the prior requirements of knowledge of $\Omega_{\rm m}$ become
more stringent.  For a $1000$ deg$^2$ experiment, only $\sigma(\Omega_{\rm
m}) \la 0.03$ is required (see Figure \ref{figpriorskaos}). 

Large-scale structure and/or supernovae constraints together with the
first-year WMAP CMB data currently deliver $\sigma(\Omega_{\rm m}) \approx
0.02$--0.04 assuming a flat Universe
\citep[e.g.][]{Sper03,Teggers04,Cole05}.  \cite{Balbi03} quote $\sigma(h)
\simeq 0.02$ as attainable with Planck (equivalent to $\sigma(\Omega_{\rm
m}) \simeq 0.02$), even allowing for uncertainty in the dark energy model. 
However, in light of Figure \ref{figpriorsgen}, this may not be sufficient
for a $10{,}000$ deg$^2$ baryon oscillations survey. 

Table \ref{tabw0w1} lists some $68\%$ confidence ranges for dark energy
parameters $(w_0,w_1)$ for a range of survey configurations and
cosmological priors, assuming a fiducial model $(-1,0)$. 

\begin{table*}[htbp]
\begin{center}
\begin{tabular}{cccccc}
\hline
Survey & Configuration & $\sigma(\Omega_{\rm m} h^2)$ &
$\sigma(\Omega_{\rm m})$ & $\sigma(w_0)$ & $\sigma(w_1)$ \\
\hline
spec-$z$ & (10000 deg$^2$, $0.5<z<3.5$) & 0 & 0 & 0.03 & 0.06 \\
& & WMAP & 0.01 & 0.04 & 0.07 \\
& & WMAP & 0.03 & 0.08 & 0.10 \\
& & Planck & 0.03 & 0.07 & 0.09 \\
\hline
KAOS & (1000 deg$^2$, $0.5<z<1.3$) + (400 deg$^2$, $2.5<z<3.5$) & 0 & 0 & 0.17 & 0.48 \\
& & WMAP & 0.03 & 0.27 & 0.63 \\
& & WMAP & 0.05 & 0.34 & 0.71 \\
& ($z \sim 1$) + (1000 deg$^2$, $1.5<z<2.5$) & 0 & 0 & 0.10 & 0.26 \\
\hline
SKA & (20000 deg$^2$, $0.5<z<1.5$) & 0 & 0 & 0.04 & 0.11 \\
& & Planck & 0.01 & 0.05 & 0.13 \\
& & Planck & 0.03 & 0.11 & 0.18 \\
\hline
photo-$z$ & (10000 deg$^2$, $0.5<z<3.5$, $\sigma_0=0.01$) & 0 & 0 & 0.07 & 0.19 \\
& & WMAP & 0.01 & 0.20 & 0.57 \\
& & WMAP & 0.03 & 0.30 & 0.95 \\
& (10000 deg$^2$, $0.5<z<3.5$, $\sigma_0=0.03$) & 0 & 0 & 0.12 & 0.32 \\
& & WMAP & 0.01 & 0.23 & 0.66 \\
& & WMAP & 0.03 & 0.31 & 0.95 \\
& (2000 deg$^2$, $0.5<z<3.5$, $\sigma_0=0.01$) & 0 & 0 & 0.19 & 0.51 \\
& & WMAP & 0.01 & 0.25 & 0.72 \\
& & WMAP & 0.03 & 0.31 & 0.95 \\
\hline
\end{tabular}
\end{center}
\caption{Simulated $68\%$ confidence ranges for the dark energy
  parameters $(w_0,w_1)$ for a series of future spectroscopic and
  photometric redshift surveys, assuming a fiducial cosmology
  $(-1,0)$.  A range of (Gaussian) priors on the values of
  $\Omega_{\rm m} h^2$ and $\Omega_{\rm m}$ are considered.  The WMAP
  and Planck measurement precisions of $\sigma(\Omega_{\rm m} h^2)$
  are assumed to be $0.004$ and $0.001$, respectively.  We assume
  $\Omega_{\rm k} = 0$.}
\label{tabw0w1}
\end{table*}

\begin{figure}[htbp]
\begin{center}
\epsfig{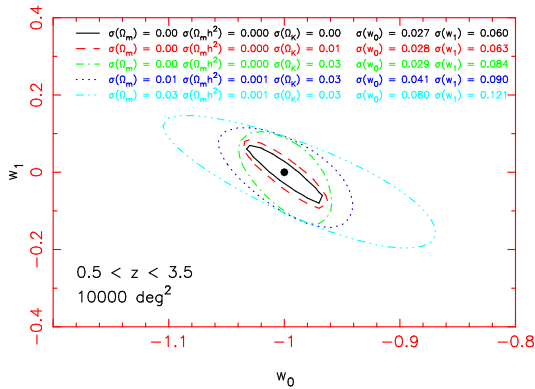}
\end{center}
\caption{Likelihood contours ($68\%$) for dark energy model parameters
  $(w_0,w_1)$ for the same (combined) survey as Figure
  \ref{figw0w1gen}, considering various different Gaussian priors upon
  the value of $\Omega_{\rm k} = 0$ with standard deviations
  $\sigma(\Omega_{\rm k})$.  We consider both holding fixed the values
  of $\Omega_{\rm m}$ and $h$, and marginalizing over these
  parameters.}
\label{figcurvgen}
\end{figure}

Throughout this paper we assume a spatially-flat ($\Omega_{\rm k} = 0$)
cosmology.  In Figure \ref{figcurvgen} we compute how the likelihood
contours in the $(w_0,w_1)$ plane for our $10{,}000$ deg$^2$ survey
degrade as our knowledge of $\Omega_{\rm k}$ weakens. We note that current
determinations of the curvature \citep[$\sigma(\Omega_{\rm k}) \la
0.02$,][]{Sper03,Dan05} are almost adequate for this proposed experiment. 
Of interest is \cite{Bern05} which noted that the the {\it combination\/} of baryon oscillations
with weak lensing constraints leads to direct breaking of degeneracies of curvature with
dark energy and allows $\Omega_{\rm k}$ to be measured without any assumptions about
the equation of state.

\subsection{Comparison with the Fisher matrix methodology}

It is worth comparing our methodology and results with the Fisher matrix
approach utilized by \cite{SE03} for the simulation of baryonic
oscillations experiments.  The input data and assumptions are not entirely
consistent in the two cases, but we can make a reasonably direct
comparison of, for example, Figure 5 in \cite{SE03} with Figure
\ref{figw0w1kaos} in this study.  In such comparisons we find that the
accuracies of determination of $(w_0,w_1)$ are consistent within a factor
of about $1.5$, with the Fisher matrix method yielding tighter contours. 

This is not surprising: the Fisher matrix method uses information from the
whole power spectrum shape (which will also be distorted in an assumed
cosmology, e.g.\ by the Alcock-Paczynski effect) whereas in our approach,
this shape is divided out and only the oscillatory information is
retained.  The bulk of the Fisher information does appear to originate
from the sinusoidal features (see Figure 5 in \cite{HH03} and Section 4.4
of \cite{SE03}).  However, the improvement in dark energy precision
resulting from fitting a full power spectrum template to the data may be
up to $50\%$ compared to our `model-independent' treatment.  Our results
should provide a robust lower limit on the accuracy, as intended. 

We also note that the Fisher approach provides the {\it minimum
  possible} errors for an unbiased estimate of a given parameter based
upon the curvature of the likelihood surface near the fiducial model.
As a result the projected error contours for any combination of two
parameters always form an ellipse (e.g. Figure~5 of \cite{SE03}).  In
our approach we explore the whole parameter space and estimate
probabilities via Monte Carlo techniques: the error contours are thus
larger and not necessarily elliptical (e.g.\ our Figure
\ref{figw0w1kaos}).

A further difference between the appearances of our Figure
\ref{figw0w1kaos} and Figure 5 in \cite{SE03} is a noticeable change in
the tilt of the principal degeneracy direction in the $(w_0,w_1)$ plane. 
The reason for this is readily identified: Seo \& Eisenstein additionally
incorporate the CMB measurement of the angular diameter distance to
recombination into their confidence plots.  We choose not to do this in
order to expose the low redshift independent cosmological constraints from
galaxy surveys and to isolate our results from the effects of any unknown
behaviour of the equation of state (which could extend beyond our $w_0,
w_1$ formalism) between $z\sim 4$ and $z\sim 1100$. 

In summary, we are encouraged by the rough agreement in values of $\Delta
w_0$ and $\Delta w_1$ between these two very different techniques.  They
represent respectively more conservative/robust and best possible dark energy
measurements from future surveys for baryon oscillations. 

\section{Detectability and accuracy of wavescale extraction}
\label{secdetect}

In this Section we take a step back from questions of dark energy and
employ our simulation tools to re-consider the fundamental question of the
detectability of the acoustic oscillations in $P(k)$ as a function of
survey size and redshift coverage. 

{\it The oscillations in the galaxy power spectrum are a fundamental
  test of the paradigm of the origin of galaxies in the fluctuations
  observed in the early Universe via the CMB.}

Detection of the oscillations would be an extremely important validation
of the paradigm.  Furthermore our standard ruler technique cannot be
confidently employed unless the sinusoidal signature in the power spectrum
can be observed with a reasonable level of significance. 

Recently, analysis of the clustering pattern of SDSS Luminous Red Galaxies
at $z = 0.35$ (a volume of $\sim 3.5 V_{\rm Sloan}$) has yielded the first
convincing detection of the acoustic signal and application of the
standard ruler \citep{Dan05}.  Although this survey does not have
sufficient redshift reach to strongly constrain dark energy models, this
result is an important validation of the technique.  Analysis of the final
database of the 2dF Galaxy Redshift Survey has also yielded some visual
evidence for baryonic oscillations \citep{Cole05}.  Here we make the
distinction between {\it detection of oscillations} and {\it detection of
a baryonic signal} $\Omega_{\rm b} \ne 0$ in the clustering pattern:
baryons produce an overall shape distortion in $P(k)$ as well as the
characteristic pattern of oscillations.  In this Section we define the
`wiggles detectability' as the confidence of rejection of a smooth `no
wiggles' model (i.e.\ the best-fitting smooth reference spectrum of step
12 in Section \ref{secsteps}).  This is a different quantity to the
`$3.4$-sigma confidence' of observing $\Omega_{\rm b} \ne 0$ reported by
\cite{Dan05}.  However, the techniques roughly agree on the measurement
accuracy of the standard ruler (see the discussion of SDSS LRGs in our
Paper I, Figure 3). 

\begin{figure}[htbp]
\begin{center}
\epsfig{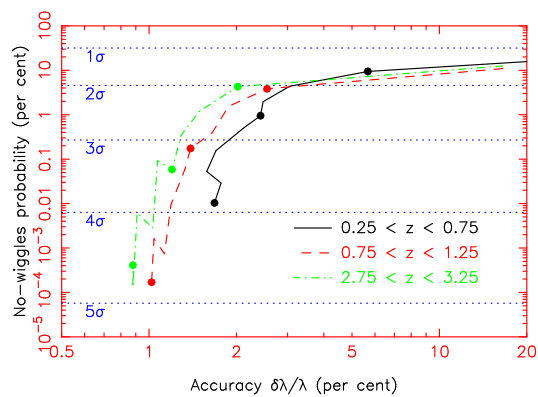}
\end{center}
\caption{Variation of the `wiggles detectability' (i.e.\ the
  significance of rejection of a no-wiggles model), and the accuracy
  with which the characteristic sinusoidal scale may be extracted,
  with survey volume.  Three surveys are considered at different
  redshifts, centred upon $z=0.5$, $z=1$ and $z=3$.  The lines join
  points separated by volume intervals of $1 V_{\rm Sloan}$; the three
  solid circles on each line denote volumes $(2, 5, 10) \, V_{\rm
    Sloan}$.}
\label{figdetect}
\end{figure}

Figure \ref{figdetect} tracks the detection significance against the
percentage accuracy of recovery of the standard ruler for surveys covering
three different redshift ranges: $0.25 < z < 0.75$, $0.75 < z < 1.25$ and
$2.75 < z < 3.25$.  The lines connect points separated by survey volume
intervals of $1 V_{\rm Sloan}$; the solid circles denote volumes $(2, 5,
10) \, V_{\rm Sloan}$. For the purposes of this Section,
an angle-averaged (isotropic) power spectrum is used rather than a power
spectrum separated into tangential and radial components.  We also only
consider the pure vacuum $\Lambda$CDM model: the detectability is
primarily driven by the cosmic volume surveyed, which is a relatively slow
function of dark energy parameters.  The detection significance is
calculated from the average over the Monte Carlo realizations of the
relative probability $P_{\rm rel}$ of the smooth `no-wiggles' model and
best-fitting `wiggles' model, where
\begin{equation}
P_{\rm rel} = \exp{[-(\chi^2_{\rm no-wig} - \chi^2_{\rm best-wig})/2]} 
\label{eqprel}
\end{equation}

We note that:

\begin{enumerate}

\item In order to obtain a significant ($3\sigma$) rejection of a
  no-wiggles model without using any power spectrum shape information,
  several $V_{\rm Sloan}$ must be surveyed.  For the surveys centred
  at $z=(0.5,1,3)$ the required volume (in units of $V_{\rm Sloan}$)
  is approximately $(7,5,5)$.  For the redshift ranges listed above,
  this corresponds to survey areas $(2300,700,500)$ deg$^2$.

\item For a fixed wiggles detection significance, the accuracy of
  recovery of the standard ruler increases with redshift.  This is due
  to the larger available baseline in $k$ at higher redshift, owing to
  the extended linear regime.  Two full oscillations are visible at $z
  = 1$; four are unveiled by a survey at $z = 3$.

\item For a fixed wavescale accuracy, the detection significance
  decreases with redshift, because the amplitude of the oscillations
  is damped with increasing $k$.  As noted above, at higher redshifts
  there are more acoustic peaks available, thus a less significant
  measurement of each individual peak may be tolerated.

\item The distribution of values of $P_{\rm rel}$ (equation
  \ref{eqprel}) over the Monte Carlo realizations is significantly
  skewed (see the discussion in \cite{BB05}).  The median value of
  $P_{\rm rel}$ represents a more confident detection than the average
  plotted in Figure \ref{figdetect}.

\end{enumerate}

\begin{figure*}[htbp]
\begin{center}
\epsfig{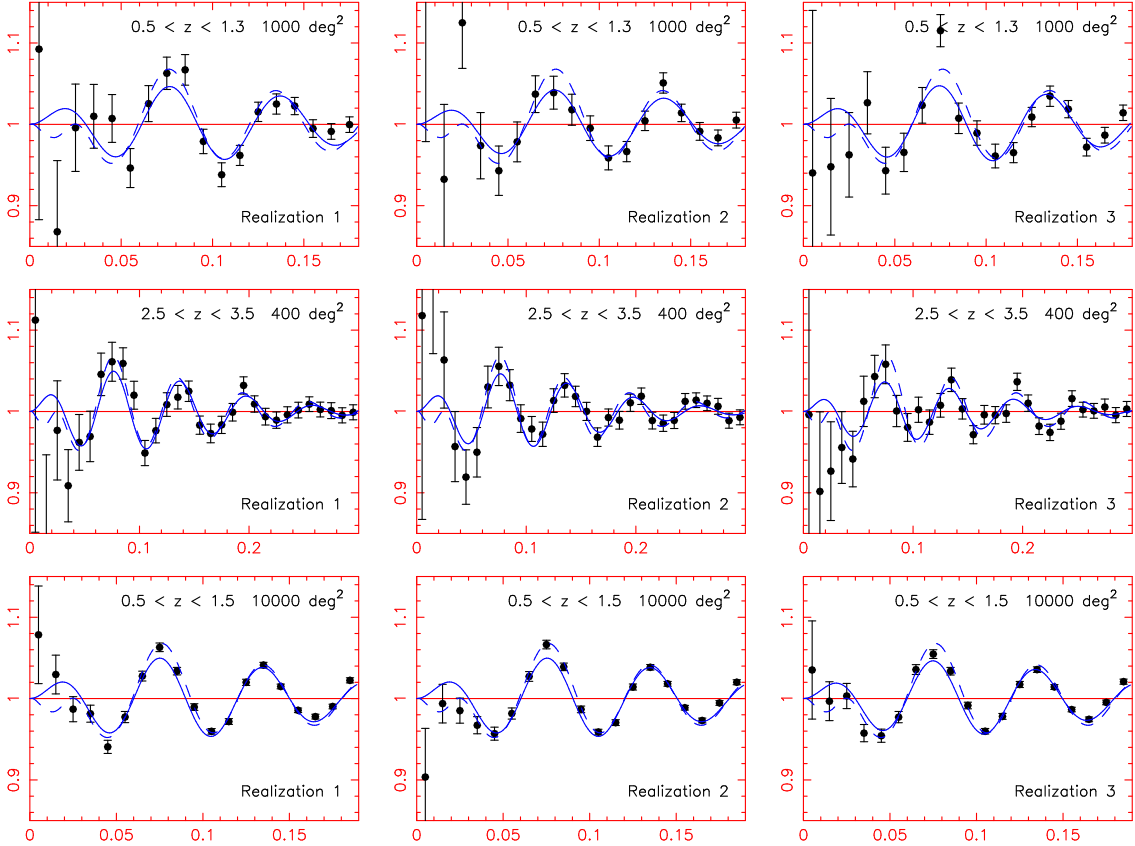}
\end{center}
\caption{Power spectrum realizations of three example high-redshift
  surveys, varying survey volume and redshift range.  In all cases, we
  plot the angle-averaged power spectrum divided by the smooth
  reference spectrum.  The dashed curve is the theoretical input
  $P(k)$ and the solid line is the best fit of our simple decaying
  sinusoidal function (Paper I, equation 3).  The $x$-axis is marked
  in units of $k$ (in $h$ Mpc$^{-1}$) and represents the extent of the
  linear regime at the redshift in question (i.e.\ $z_{\rm eff} =
  (z_{\rm min} + z_{\rm max})/2$).  The rows of the Figure represent
  surveys with geometries ($0.5 < z < 1.3$, $1000$ deg$^2$), ($2.5 < z
  < 3.5$, $400$ deg$^2$) and ($0.5 < z < 1.5$, $10{,}000$ deg$^2$);
  the columns display the first three Monte Carlo realizations in each
  case.}
\label{figpkspec}
\end{figure*}

Figure \ref{figpkspec} displays some Monte Carlo power spectrum
realizations of three surveys: ($0.5 < z < 1.3$, $1000$ deg$^2$), ($2.5 <
z < 3.5$, $400$ deg$^2$) and ($0.5 < z < 1.5$, $10{,}000$ deg$^2$).  The
total volumes mapped in units of $V_{\rm Sloan}$ are $(10,8,133)$,
respectively.  The total numbers of galaxies observed in each survey (to
ensure $n \times P = 3$) are $(6,2,85) \times 10^6$ (assuming linear bias
$b=3$ for the $z \sim 3$ survey and $b=1$ otherwise).  The first two
redshift surveys could be performed by a next-generation wide-field
optical spectrograph such as the KAOS instrument proposed for the Gemini
telescopes \citep[][ {\tt http://www.noao.edu/kaos/}]{KAOS}.  The third
survey is possible in 6 months using the Square Kilometre Array to detect
HI emission line galaxies \citep{AR04} or from a space mission (see Section~\ref{sec:space}).

\section{Dark energy measurements from realistic spectroscopic redshift
surveys}
\label{secmeas}

We now consider the prospects of performing these experiments with
realistic galaxy redshift surveys.  As noted above and in Paper I, such
surveys must cover a minimum of several hundred deg$^2$ at high redshift,
cataloguing at least several hundred thousand galaxies, in order to obtain
significant constraints upon the dark energy model. 

\subsection{Existing surveys}

These requirements are orders of magnitude greater than what has been
achieved to date.  Some existing surveys of high-redshift galaxies are the
Canada-France Redshift Survey \citep[CFRS; a few hundred galaxies covering
$\sim 0.1$ deg$^2$ to $z \approx 1.3$;][]{Lilly95}, the survey of $z \sim
3$ Lyman-break galaxies by \cite{Stei03} (roughly a thousand galaxies
across a total area $\approx 0.4$ deg$^2$).  Most other high redshift
spectroscopic surveys \citep[e.g.][]{GDDS,K20,Stei04} cover equally small
areas $\la 1$ deg$^2$. 

Some larger surveys are in progress: the DEEP2 project \citep{Davis03},
using the DEIMOS spectrograph on the Keck telescope, aims to obtain
spectra for $60{,}000$ galaxies ($3.5$ deg$^2$, $0.7 < z < 1.4$); the
VIRMOS redshift survey \citep{LeFev03}, using the VIMOS spectrograph at
the VLT, will map $150{,}000$ redshifts over $16$ deg$^2$ (considering the
largest-area component of each).  Neither of these existing projects comes
close to meeting our goals, primarily due to the limitations of existing
instrumentation.  The spectrographs used to perform these surveys have
typical fields-of-view (FOV) of diameter $10-20'$ and are unable to cover
hundreds of deg$^2$ in a reasonable survey duration. 

\subsection{New ground-based approaches (optical/IR)}

Some proposed new optical instrumentation addresses this difficulty,
permitting spectroscopic exposures over considerably larger FOVs using the
8-metre telescopes that are required to obtain spectra of sufficient
quality at these redshift depths.  For example, the KAOS project for the
Gemini telescopes \citep[][ {\tt http://www.noao.edu/kaos/}]{KAOS} is a
proposal for a $1.5$ deg diameter FOV, 4000 fibre-fed optical
spectrograph.  There are two proposed redshift surveys: $900{,}000$ ($0.5
< z < 1.3$) galaxies over $1000$ deg$^2$, and $600{,}000$ ($2.5 < z <
3.5$) galaxies across $400$ deg$^2$. These surveys would together take
$\sim 170$ clear nights using the 8-metre Gemini telescope with realistic
exposure times computed for the KAOS instrument sensitivity.  The redshift
ranges are driven by the strong spectral features available for redshift
measurement in optical wavebands in a relatively short exposure time.  The
$z \sim 1$ range is cut off at $z = 1.3$ by the [OII] emission line and
the calcium H \& K lines shifting to red/infrared wavelengths $>0.9$
\micron\ where the airglow is severe and conventional CCD detectors have
low efficiency; the $z \sim 3$ component is driven by observing Ly$\alpha$
in the blue part of the optical range. 

\begin{figure}[htbp]
\begin{center}
\epsfig{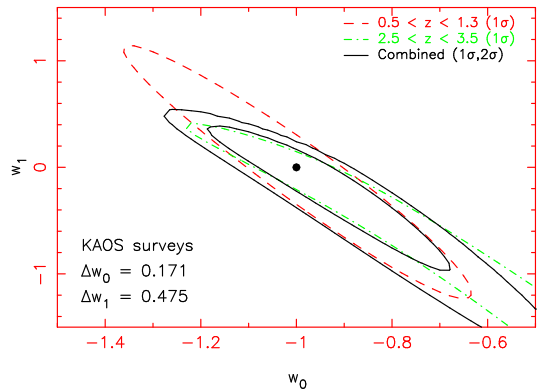}
\end{center}
\caption{Measurement of a dark energy model $w(z) = w_0 + w_1 z$ using
  a simulated survey with the KAOS spectrograph consisting of two
  components, $0.5 < z < 1.3$ ($1000$ deg$^2$) and $2.5 < z < 3.5$
  ($400$ deg$^2$).  Constraints are displayed for each redshift
  component ($68\%$), and for both surveys combined ($68\%,95\%$).
  Note that the likelihoods are generated over a much wider
  $(w_0,w_1)$ space than displayed in the Figure.  The solid circle
  denotes the fiducial cosmology, $(w_0,w_1) = (-1,0)$.  There is a
  significant degeneracy between $w_0$ and $w_1$, particularly for the
  $z \sim 3$ survey if $w_1 < 0$, owing to the lack of sensitivity of
  $H(z = 3)$ to dark energy in this case.  For this plot, perfect
  knowledge of $\Omega_{\rm m}$ ($= 0.3$), $h$ ($= 0.7$) and
  $\Omega_{\rm k}$ ($= 0$) is assumed.  The standard deviations quoted
  for $\Delta w_0$ and $\Delta w_1$ (i.e.\ half the interval between
  the $16^{\rm th}$ and the $84^{\rm th}$ percentiles) result from
  marginalizing over the other parameter (i.e.\ $w_1$ and $w_0$,
  respectively).}
\label{figw0w1kaos}
\end{figure}

The $w(z)$ measurements resulting from the proposed KAOS surveys, computed
using the methodology of Section \ref{secmeth}, are displayed in Figure
\ref{figw0w1kaos} (see also Tables \ref{tabxdx} and \ref{tabw0w1}).  We
show both the $z \sim 1$ and $z \sim 3$ contributions separately, and the
joint constraint.  We assume linear bias factors $b = (1,3)$ for the $z
\sim (1,3)$ simulations, respectively.  The measurement precision of the
dark energy parameters is $\Delta w_0 \approx 0.2$ and $\Delta w_1 \approx
0.4$, significantly better than current supernovae constraints
\citep{Riess04}.  In statistical terms the KAOS performance is somewhat
poorer than that projected for the proposed SNAP mission \citep{Ald02},
but the acoustic oscillations method is significantly less sensitive to
errors of a systematic nature.  Figure \ref{figw0w1kaos} illustrates that
models with $w_1 < 0$ are harder to exclude owing to the diminishing
sensitivity of cosmic distances to dark energy in this region of parameter
space. 

We note that the KAOS measurement of $H(z \approx 3)$ from the radial
component of the $z \sim 3$ power spectrum provides little information
about the dark energy model if we are perturbing around the cosmological
constant: at $z \sim 3$, the dynamics of the Universe are entirely
governed by the value of $\Omega_{\rm m} h^2$, and $H(z)$ is almost
independent of $(w_0,w_1)$.  However, the value of $x(z \approx 3)$
inferred from the tangential component of the $z \sim 3$ power spectrum
does depend on dark energy, because $x(z=3)$ is an integral of $dx/dz =
c/H(z)$ from $z = 0$ to $z = 3$, which is influenced by dark energy at
lower redshifts.  The $z \sim 3$ constraint thus reduces to a significant
degeneracy between $w_0$ and $w_1$, as observed in Figure
\ref{figw0w1kaos}, although models with $w_1 > 0$ can still be ruled out
by this redshift component.  The degeneracy is less severe for the $z \sim
1$ component owing to the availability of both $H(z)$ and $x(z)$
information. 

\begin{figure}[htbp]
\begin{center}
\epsfig{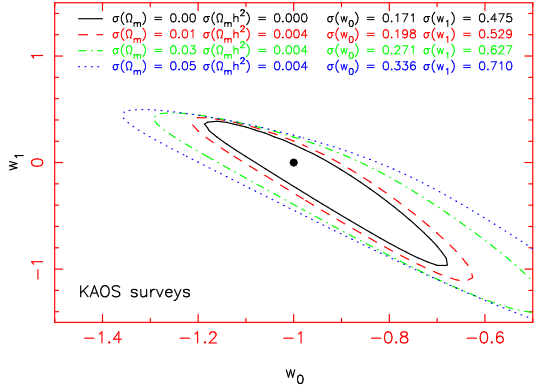}
\end{center}
\caption{Likelihoods contours ($68\%$) of dark energy parameters
  $(w_0,w_1)$ for the same KAOS redshift surveys as Figure
  \ref{figw0w1kaos}, marginalizing over the cosmological parameters
  $(\Omega_{\rm m},\Omega_{\rm m} h^2)$.  We assume a WMAP Gaussian
  prior upon $\Omega_{\rm m} h^2$ (as in Figure \ref{figpriorsgen})
  and consider a range of different Gaussian priors for $\Omega_{\rm
    m}$.  In each case, standard deviations are quoted for $w_0$ and
  $w_1$ as before.  We assume $\Omega_{\rm k} = 0$.}
\label{figpriorskaos}
\end{figure}

Figure \ref{figw0w1kaos} assumes that we have perfect prior knowledge of
$\Omega_{\rm m}$ ($= 0.3$) and $h$ ($= 0.7$).  Figure \ref{figpriorskaos}
relaxes this assumption, illustrating the effect of including Gaussian
priors upon $\Omega_{\rm m}$ and $\Omega_{\rm m} h^2$ of various widths. 

A drawback of this survey design is the absence of the redshift range $1.5
< z < 2.5$, sometimes called the `redshift desert'.  There are no strong
emission lines accessible to optical spectrographs in this interval:
existing surveys of this region have required very long exposure times to
secure spectra \citep{GDDS}, but this could be remedied by near-IR or
near-UV spectroscopy of bright, star-forming galaxies \citep{Stei04}.  We now consider
the usefulness of these additional observations as regards measuring dark
energy, along with the observational practicalities. 

\begin{figure}[htbp]
\begin{center}
\epsfig{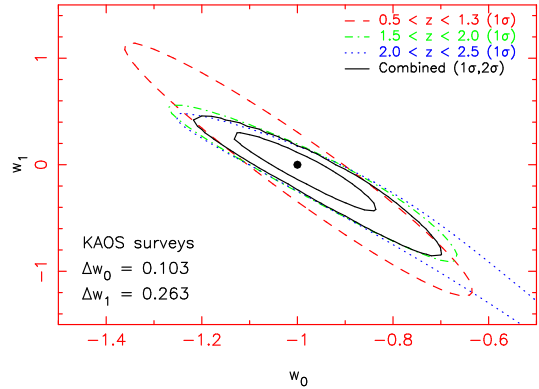}
\end{center}
\caption{Dark energy measurements resulting from $1000$ deg$^2$
  surveys spanning the redshift ranges ($0.5 < z < 1.3$) and ($1.5 < z
  < 2.5$).  We assume perfect prior knowledge of $(\Omega_{\rm
    m},h,\Omega_{\rm k})$.}
\label{figw0w1kaos2}
\end{figure}

\begin{figure}[htbp]
\begin{center}
\epsfig{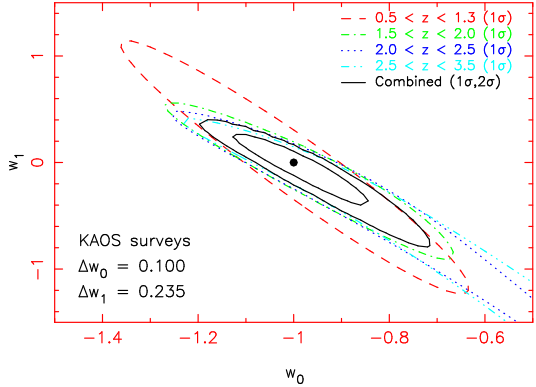}
\end{center}
\caption{Dark energy measurements resulting from $1000$ deg$^2$
  surveys spanning the redshift ranges ($0.5 < z < 1.3$) and ($1.5 < z
  < 2.5$), together with a $400$ deg$^2$ component covering ($2.5 < z
  < 3.5$).  We assume perfect prior knowledge of $(\Omega_{\rm
    m},h,\Omega_{\rm k})$.}
\label{figw0w1kaos3}
\end{figure}

First, we investigate the effect of this redshift range upon measurements
of the dark energy model $w(z) = w_0 + w_1 z$ (assuming precise prior
knowledge of $\Omega_{\rm m}$ and $h$ for the purposes of this comparison;
Figure \ref{figpriorskaos} indicates how accurately these parameters must
be known in order that their uncertainty is not limiting).  In Figure
\ref{figw0w1kaos2}, we remove the $z = 3$ component of the proposed KAOS
experiment and extend the lower-redshift $1000$ deg$^2$ survey across the
redshift range $1.5 < z < 2.5$, divided into two independent slices of
width $\Delta z = 0.5$.  The likelihood constraints in $(w_0,w_1)$ space
tighten appreciably, {\it by a further factor of two}, principally due to
the $1.5 < z < 2.0$ component, for which $H(z)$ still yields useful
information about $(w_0,w_1)$. In Figure \ref{figw0w1kaos3} we add back in
the $z = 3$ data; the dark energy measurements do not significantly
improve. We conclude that coverage of the redshift desert would be highly
desirable if it could be achieved. Simply increasing the area of the
$z\sim 3$ component does not help nearly as much ($\Delta w_1$ is improved
by $\sim 25\%$ for 1000 deg$^2$ at $z\sim 3$).  This simplistic analysis
is of course no substitute for a proper survey optimization assuming fixed
total time or cost. 

We note, however, that there are other persuasive scientific reasons to
include a $z = 3$ survey component \citep{Dan02}, amongst them: 

\begin{enumerate}

\item A galaxy redshift survey at $z = 3$ unveils the linear power
  spectrum down to unprecedentedly small scales $k \approx 0.5 \, h$
  Mpc$^{-1}$, measuring linear structure modes that cannot be accessed
  using the CMB.

\item If dark energy is insignificant at $z = 3$, then measurement of
  the acoustic oscillations in such a redshift slice enables the
  standard ruler to be calibrated in a manner independent of the CMB.

\item Our present analysis assumes that the fiducial dark energy model
  is a cosmological constant, $(w_0,w_1) = (-1,0)$.  Based on current
  data, we have very little information about the value of $w_1$ (for
  the latest supernova analysis of \cite{Riess04}, $\sigma(w_1)
  \approx 0.9$).  Should $w_1 \ne 0$, the influence of $w(z)$ upon
  higher-redshift dynamics could become more significant. A general
  redshift survey optimization, which is beyond the scope of this
  paper, should address the range of $w(z)$ to be explored
  \citep{Bassett05}.

\end{enumerate}

\begin{figure}[htbp]
\begin{center}
\epsfig{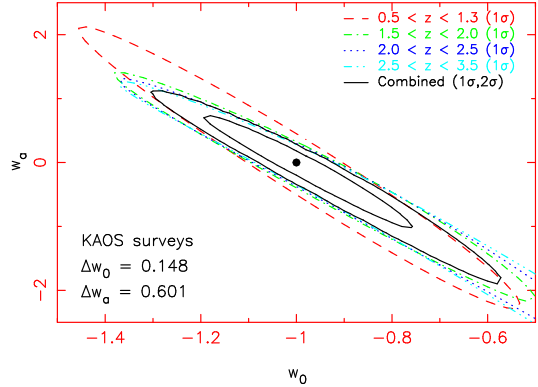}
\end{center}
\caption{Dark energy measurements resulting from the same surveys as
  Figure \ref{figw0w1kaos3}, using the alternative parameterization
  $w(z) = w_0 + w_a(1-a)$.  We assume perfect prior knowledge of
  $(\Omega_{\rm m},h,\Omega_{\rm k})$.  Note that the likelihoods are
  generated over a much wider $(w_0,w_a)$ space than displayed in the
  Figure.}
\label{figw0wa}
\end{figure}

Figure \ref{figw0wa} considers a different parameterization for dark
energy $w(z) = w_0 + w_a (1-a)$ $=w_0 + w_a/(1+z)$ 
(Linder 2002), where $a=(1+z)^{-1}$ is the usual
cosmological scale factor, which encapsulates a more physically realistic
behaviour at high redshifts $z \gtrsim 1$ in comparison to $w(z) = w_0 +
w_1 z$.  The rate of change of $w$ with redshift in the two models is
$dw/dz = w_1 = w_a/(1+z)^2 < w_a$ hence for a given survey we expect the
size of the error for $w_a$ to exceed that for $w_1$.  Figure
\ref{figw0wa} illustrates this for the same survey configuration as Figure
\ref{figw0w1kaos3}. 

\begin{figure}[htbp]
\begin{center}
\epsfig{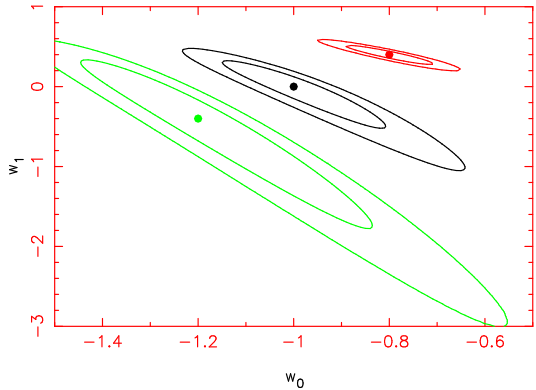}
\end{center}
\caption{The tightness of the $(68\%,95\%)$ likelihood contours in the
  $(w_0,w_1)$ plane is a strong function of the fiducial cosmological
  model.  In this Figure we illustrate, for three different fiducial
  models $(w_0,w_1) = (-1,0)$, $(-0.8,0.4)$ and $(-1.2,-0.4)$, dark
  energy measurements resulting from $1000$ deg$^2$ surveys spanning
  the redshift ranges ($0.5 < z < 1.3$) and ($1.5 < z < 2.0$),
  together with a $400$ deg$^2$ component covering ($2.5 < z < 3.5$).
  We assume perfect prior knowledge of $(\Omega_{\rm m},h,\Omega_{\rm
    k})$.}
\label{figfid}
\end{figure}

Figure \ref{figfid} uses the KAOS surveys plus a $1.5 < z < 2.0$ extension
to illustrate how the tightness of the likelihood contours in the
$(w_0,w_1)$ plane is a strong function of the fiducial dark energy
parameters, as discussed in Section \ref{secstream}.  We computed the
linear growth factor for non-$\Lambda$CDM models by solving the
appropriate second-order differential equation \citep[e.g.][]{LinJen}.  If
$w_1 > 0$ then dark energy is more significant at higher redshifts and the
model parameters can be constrained more tightly, despite both the
resulting decrease in the available cosmic volume in a given redshift
range and the movement of the linear/non-linear transition to larger
scales (smaller $k$).  Note that our cut-off to the evolution of $w(z) =
w_0 + w_1 z$ at $z_{\rm cut} = 2$ ensures that dark energy does not
dominate at high redshift for our model with $w_1 > 0$. 

Next, we consider the practicalities of observing galaxies in the redshift
range $1.5 < z < 2.5$ (the so-called `optical redshift desert'). 
Considering near infra-red wavebands first: the H$\alpha$ 6563\AA\ line is
accessible in the 1--2$\mu$m band over the interval $0.5 < z < 2$.  This
regime is the non-thermal infra-red, in which the sky is sufficiently dark
to permit high-redshift spectroscopy (e.g.\ the emission-line observations
of \cite{Glz99} and \cite{Pett98}).  The critical opportunity offered here
is that the H$\alpha$ emission line is potentially very bright at
redshifts $z \gtrsim 1$, owing to the steep evolution in the
star-formation rate of galaxies in the Universe over $0 < z \lesssim 1$
\citep{Hop00}. 

Let us consider a redshift slice $1.5 < z < 1.7$.  Our requirement for
shot noise sampling ($n \times P = 3$) translates into a required surface
density $(3900/b^2)$ deg$^{-2}$ within this slab (Figure \ref{fignp}),
where $b$ is the linear bias factor of the surveyed galaxies.  The
luminosity function of H$\alpha$ emitters at $z \gtrsim 1$ has been
reasonably well determined from NICMOS slitless grism surveys using the
Hubble Space Telescope \citep{Hop00,Yan99}.  Using the Hopkins et al.\
luminosity function, we find that we need to reach a line flux limit of
$1.1 \times 10^{-16}$ ergs cm$^{-2}$ s$^{-1}$ in order to reach the
required surface density (assuming $b=1$).  The $z=1$ H$\alpha$ luminosity
function has apparently evolved strongly in comparison with $z=0$
\citep{Gal95} but the measurements are fairly robust: we can double-check
the luminosity function determinations by simply counting objects in the
NICMOS surveys above our required line flux limit.  In the Hopkins,
Connolly \& Szalay sample there are 4 galaxies in the redshift range $1.5
< z < 1.7$ above this flux limit, spread over 4.4 arcmin$^2$, yielding a
surface density $\simeq$ 3300 deg$^{-2}$.  The H$\alpha$ identifications
are also very reliable: the Yan et al.  sample was observed in optical
wavebands by \cite{Hicks02}, confirming $\ge 75\%$ of the H$\alpha$
identifications via associated [OII] emission at the same redshift. This
agrees with expectations: analytic models of evolving line emission show
that H$\alpha$ should dominate at these flux levels at 1-2\micron\ over
other lines. 

This is encouraging, because these bright lines are accessible in
relatively modest exposures.  Let us assume an 8-metre telescope, a $25\%$
efficient $R = 4000$ near-IR spectrograph and detector (with a dispersion
of 5\AA\ per arcsec), and consider an object observed in an aperture of
size 0.8 arcsec $\times$ 0.8 arcsec, covering $2\times 2$ pixels and
containing half the light (the Yan et al.\ objects have compact half-light
radii of 0.2--0.7 arcsec).  The signal-to-noise ratio is determined by
detector readout noise, sky background and dark current.  We will assume a
readout noise of 4 electrons (this is the stipulated requirement of the
James Webb Space Telescope (JWST) detectors) and an inter-OH sky
background of 1.2 photons s$^{-1}$ nm$^{-1}$ arcsec$^{-2}$ m$^{-2}$ in the
H-band (the OH night sky line forest is well-resolved at $R=4000$) which
is a typical value measured at Gemini observatory\footnote{See
  \url{http://www.gemini.edu/sciops/ObsProcess/obsConstraints/\\
    ocSkyBackground.html}}.  We will neglect dark current, which is
equivalent to assuming that this dark current is much lower than the
sky counts.  Given these assumptions, our H$\alpha$ flux limit at $z =
1.6$ corresponds to a signal-to-noise ratio of 20 in a 600 second
integration, and we find the observation is background limited for any
readout noise $<$ 15 electrons.

This exposure time is encouragingly short, and implies that using a 1
degree FOV spectrograph, one could survey $1000$ deg$^2$ in only 20
nights.  Admittedly our instrument specification is optimistic, especially
for readout noise; however, it can be relaxed considerably whilst still
achieving exposure times $< 1$ hour. Assuming a fibre spectrograph we
estimate that observing 3900 objects simultaneously at this resolution
would only require 2--3 detector arrays of size 4096$\times$4096 to cover
the $H$-band. Obviously more objects or broader wavelength coverage would
require more detectors or time. Finally we note that, at least in
principle, the exposure times are sufficiently short that one could
imagine performing such a survey on a smaller-aperture (4-metre class)
telescope. 

The potential problem with the approach outlined above is that it is not
known `a priori' which of the galaxies identified in deep images will be
H$\alpha$ bright, or the redshifts of these galaxies. Possibly these data
could be successfully predicted from other information (e.g.\ broad-band
colours or sub-mm/radio fluxes), but this may be unreliable.  Targetting
fainter H$\alpha$ fluxes resulting from realistic target selections would
obviously require longer exposure times.  Further work on this problem is
required in order to determine the distinguishing properties of known
H$\alpha$-bright galaxies.  

A second potential difficulty is the effect of the night-sky OH emission
lines in potentially making inaccessible certain redshift ranges, in a
complex pattern. Since these redshift ranges are very narrow, the result
is the removal of a series of redshift spikes at known locations in the
radial window function.  In order to assess the likely consequences, we
manufactured a synthetic $n(z)$ possessing narrow gaps where no galaxies
could be observed, i.e.\ when the redshifted H$\alpha$ emission line
coincided with an OH line or landed in the water-absorption hole between
the J and H bands.  We assumed a spectrograph operating at a resolution $R
= 4000$, which implied that $68\%$ of the $1.1 < z < 1.7$ redshift
interval was accessible.  With our simulation tools we then recovered
$P(k)$ using this $n(z)$ (employing an FFT with sufficient gridding in the
radial direction to resolve these narrow spikes).  The principal effect of
the window function is to damp the amplitude of the acoustic oscillations
slightly in the radial direction, leaving the tangential modes largely
unaffected.  The fractional error ($\Delta k_A/k_A$) with which the
acoustic scale is recovered is increased by no more than $25\%$, which is
mostly due to the smaller effective survey volume owing to the absence of
many thin redshift shells.  We conclude that the OH lines are not a factor
which will significantly hamper ground-based surveys of the acoustic
peaks. 

Bright star-forming galaxies can alternatively be observed by targetting
the [OII] 3727\AA\ emission line in optical wavebands. High-resistivity
CCD detectors under development can maintain quantum efficiency out to
1\micron\ \citep{CCD} corresponding to $z=1.7$.  The typical
H$\alpha$:[OII] intensity is 2:1; high spectral resolution is again
required to observe between the OH night sky lines, in which case the
exposure times would be short if one could pre-select the [OII]-bright
population. This could be more problematic than with H$\alpha$ as there is
considerable extra scatter introduced in to the line ratio [OII]$/$H$\alpha$ due to variations
in metallicity and extinction \citep{JFF01}. We also note that [OII] is
accessible in the non-thermal IR up to $z=5$ and therefore is a potential
probe of high-redshift acoustic oscillations. 

The high star-formation rate at earlier cosmic epochs also implies a
considerable boost in the rest-frame UV luminosity of galaxies.
High-altitude sites such as Mauna Kea have an atmospheric cutoff further
into the near-UV, which can be exploited by blue-optimized spectrographs. 
For example, \cite{Stei04} have obtained spectra of star-forming galaxies
over $1.4<z<2.5$ using exposure times of only a few hours at the 10-metre
Keck telescope, reaching a lowest observed-frame wavelength of 3200\AA;
this corresponds to Ly$\alpha$ at $z = 1.6$, or CIV at $z = 1.1$.  The
spectral region between Ly$\alpha$ and CIV is rich in interstellar lines
and ripe for determination of accurate redshifts.  It is possible that a
UV approach may be superior to a near-IR approach targetting H$\alpha$:
the observed surface density of Steidel et al. sample is sufficiently high
for our requirements, but we note that a UV-optimized design would
probably require a wide-field slit spectrograph because conventional
fibres considerably attenuate the UV light for long runs ($>20$m). 

\subsection{New ground-based approaches (radio)}

Next-generation radio interferometer arrays, such as the proposed Square
Kilometre Array (SKA; {\tt http://www.skatelescope.org}; planned to
commence operation in about 2015), will have sufficient sensitivity to
detect the HI (21cm) transition of neutral hydrogen at cosmological
distances that are almost entirely inaccessible to current radio
instrumentation.  This will provide a very powerful means of performing a
large-scale redshift survey: once an HI emission galaxy has been located
on the sky, the observed wavelength of the emission line automatically
provides an accurate redshift. 

The key advantage offered by a radio telescope is that it may be designed
with an instantaneous FOV exceeding $100$ deg$^2$ (at $1.4$ GHz), vastly
surpassing the possibilities of optical spectrographs. Equipped with a
bandwidth of many $100$ MHz, such an instrument could map out the cosmic
web (probed by neutral hydrogen) at an astonishing rate: the SKA, if
designed with a large enough FOV, could locate $\sim 10^9$ HI galaxies to
redshift $z \approx 1.5$ over the whole visible sky in a timescale of
$\sim 1$ year \citep{AR04,BABR04}.  Deeper pointings could probe the HI
distribution to $z \sim 3$ over smaller solid angles.  A caveat is that
the HI mass function of galaxies has been determined locally
\citep{Zwaan03} but is currently very poorly constrained at high redshift. 
However, for a range of reasonable models, the number densities required
to render shot noise negligible may be attained for HI mass limits {\it
larger} than the break in the mass function \citep{AR04}.  Another
requirement of interferometer design is that a significant fraction of the
collecting area must reside in a core of diameter a few km, to deliver the
necessary surface brightness sensitivity for extended 21cm sources.  Sharp
angular resolution is not a pre-requisite, assuming that the observed
galaxies are not confused. 

\begin{figure}[htbp]
\begin{center}
\epsfig{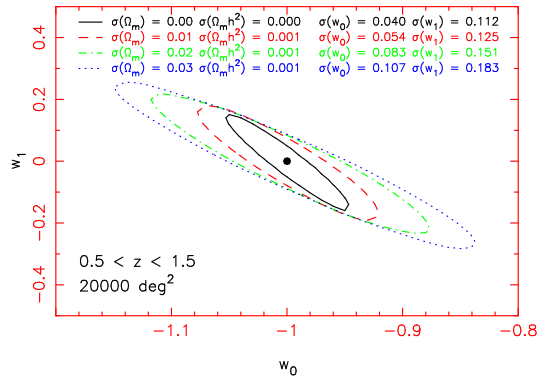}
\end{center}
\caption{Likelihoods ($68\%$) of dark energy parameters $(w_0,w_1)$
  for a future 21cm radio survey, using the Square Kilometre Array to
  map HI emission galaxies over $20{,}000$ deg$^2$, covering a
  redshift range $0.5 < z < 1.5$.  Such a survey is possible in a
  timescale of less than 1 year if the SKA is designed with a
  sufficiently large FOV ($\sim 100$ deg$^2$ at $1.4$ GHz) and
  bandwidth ($\sim 200$ MHz).  We marginalize over the cosmological
  parameters $(\Omega_{\rm m}, \Omega_{\rm m} h^2)$ using a range of
  different Gaussian priors.  Our assumed prior for $\Omega_{\rm m}
  h^2$ is representative of that obtained by the Planck satellite.  We
  assume $\Omega_{\rm k} = 0$.}
\label{figw0w1ska}
\end{figure}

Figure \ref{figw0w1ska} displays measurements of the dark energy model
resulting from a $20{,}000$ deg$^2$ neutral hydrogen survey over the
redshift range $0.5 < z < 1.5$, analyzing acoustic peaks in redshift
slices of width $\Delta z = 0.2$.  We note that a smaller 21cm survey
could be performed in the nearer future by SKA prototypes such as the
HYFAR proposal \citep{Bun03}, which may cover several thousand deg$^2$
over a narrower bandwidth (corresponding to $0.8 < z < 1.2$). 

\subsection{Space-based approaches} \label{sec:space}

An interesting alternative approach is to use a space-based dispersive but
slitless survey to pick out emission-line objects directly over a broad
redshift range.  In space, the 1--2$\mu$m background is $1000$ times lower
(in a broad band) than that observed from the ground \citep[based upon the JWST mission
background simulator at distance 3 au from the Sun;][]{Petro02}; in the
background-limited regime this gain is equivalent to a $1000$-fold
increase in collecting area.  The dispersing element could be either a
large objective prism or a grism. The NICMOS surveys mentioned above
already demonstrate that this technique is possible, but they lack
somewhat in FOV and spectral resolution compared to what is desirable.

As an illustrative example, let us consider a 0.5-metre space telescope
with a $60\%$ overall system efficiency working in slitless dispersed
imaging mode, again targeting $0.5''$ galaxies at redshift $z = 1.6$.  We
will assume the JWST-like background and that the slitless spectra are
de-limited by a 2000\AA-wide blocking filter. Just considering the sky
background, the signal-to-noise ratio in a 1800s exposure is 5 for our
canonical flux limit of $1.1 \times 10^{-16}$ ergs cm$^{-2}$ s$^{-1}$. 
This signal-to-noise ratio is independent of spectral resolution for
unresolved lines much brighter than the continuum (the spectral resolution
must of course be sufficient for determining accurate redshifts). 
Highly-dispersive IR materials such as silicon could enable an objective
prism approach with $R = 500$, using a simple prime focus imaging system. 
Such a satellite, if equipped with a 1 deg FOV and sensitive over
1--2\micron, could perform a spectroscopic survey over the redshift
interval $0.5 < z < 2$ (using H$\alpha$), covering an area of $10{,}000$
deg$^2$ in a 4 year mission, obtaining dark energy constraints very
similar to those presented in Figure \ref{figw0w1gen}.  Accessing higher
redshifts would be possible by either extending the wavelength range or
going fainter with the [OII] line, in both cases requiring a larger
diameter mirror.  Ambiguous line identification is a potential problem,
this could be remedied by cross-matching with a photometric-redshift
imaging survey. We refer the reader to \cite{BOP} for a more detailed
discussion of such a dedicated `Baryonic Oscillation Probe'.

\section{Dark energy measurements from realistic photometric redshift
surveys}
\label{secphoto}

We next explore the potential of surveys based on {\it photometric
redshifts} for detecting the acoustic oscillations and placing constraints
upon any cosmic evolution of the dark energy equation-of-state $w(z)$. 
For a detailed treatment of the significance of `wiggles detection' and
the accuracy of measurement of the standard ruler with photometric
redshift surveys, we refer the reader to \cite{BB05}.  Here we present a
summary and discuss the consequences for dark energy measurement. 

A photometric redshift (`photo-z') is obtained from multi-colour
photometry of a galaxy: the object is imaged in several broad-band
filters, ranging from the UV to the near-IR, producing a rough spectral
energy distribution (SED).  This observed SED is then fitted by model
galaxy SEDs as a function of redshift to construct a likelihood
distribution with redshift; the peak of the likelihood function indicates
the best-fitting redshift.  A classic example of this approach is the
analysis of the Hubble Deep Field North \citep{FLY99}. 

There are obviously many options concerning the number and width of filter
bands, and their placement in the UV-NIR range.  Generally at least five
broad bands are used, and IR coverage is essential for constraining
galaxies with $1.2 < z < 2.2$ \citep{BMP00}.  Some approaches have used as
many as 17 broad $+$ narrow-band filters \citep[e.g.][]{COMBO17}.  Many
different techniques have been proposed for deriving the photo-z
\citep[e.g.][]{Cs00,leB02,CL04}.  The important question for our study is:
what is the accuracy of the photo-z estimates? 

These errors can be divided into two main types.  First, there is the
random statistical error due to noise in the flux estimates and to the
coarseness of the SED. Typically this is specified by a parameter
$\sigma_0$ where
\begin{equation}
\sigma_0 = {\sigma_z \over (1+z) } \approx \hbox{constant}
\end{equation}
and $\sigma_z$ is the standard deviation of the redshift $z$.  We note
that $\sigma_0$ is approximately constant because it is proportional
to the spectral resolution $\lambda/\Delta\lambda$ of the set of
filters.  \cite{Chen03} obtained $\sigma_0 = 0.08$; the COMBO17 survey
achieved $\sigma_0 = 0.03$.  In a theoretical study, \cite{Bud01}
demonstrated that an optimized filter set produced results within the
range $\sigma_0 = 0.02 - 0.05$, depending on the shape of the SED.  In
general, redder galaxies deliver more accurate photo-z's because the
model colours change faster with redshift.

The second source of photo-z error is the possibility of getting the
redshift grossly wrong, either because the set of colours permit more than
one redshift solution, or because the model SEDs are not sufficiently
representative of real galaxies.  Different authors disagree about the
magnitude of this effect, which depends on the specific filter sets,
photometric accuracy, spectroscopic calibration and photo-z methods used. 
A useful theoretical discussion is given in \cite{BMP00}.  \cite{BB05}
analyze various `realistic' redshift error distributions including
outliers and systematic offsets. 

For our purposes we will ignore systematic errors and parameterize photo-z
performance using the value of $\sigma_0$ alone.  A realistic survey will
contain additional systematic redshift errors, thus we will obtain lower
limits.  We will also assume that $\sigma_0$ is a constant, whereas for a
realistic survey it will depend somewhat on redshift and galaxy type. 

What is the effect of a statistical redshift error ($\sigma_0$) on the
measured power spectrum?  This is fairly easy to estimate analytically. As
discussed in Paper I \cite[see also][Section~4.5]{SE03} this photo-z error
represents a radial smearing of galaxy positions.  For example, $\sigma_0
= 0.03$ for a $z = 1$ galaxy corresponds to an error $\sigma_x \approx 100
\, h^{-1}$ Mpc in the radial co-moving coordinate.  Since smoothing (i.e.\
convolution) by a Gaussian function in real space is equivalent to
multiplication by a Gaussian function in Fourier space, we can model the
photo-z effect as a multiplicative damping of the 3D power spectrum
$P(k_x, k_y, k_z)$ by a term $\exp{(-k_x^2 \sigma_x^2)}$ (where we choose
the $x$-axis as the radial direction).  In the following, we implement a
{\it flat-sky approximation} and presume that there is no tangential
effect (i.e.\ parallel to the $(y,z)$-plane). 

\begin{figure*}[htbp]
\begin{center}
\epsfig{file=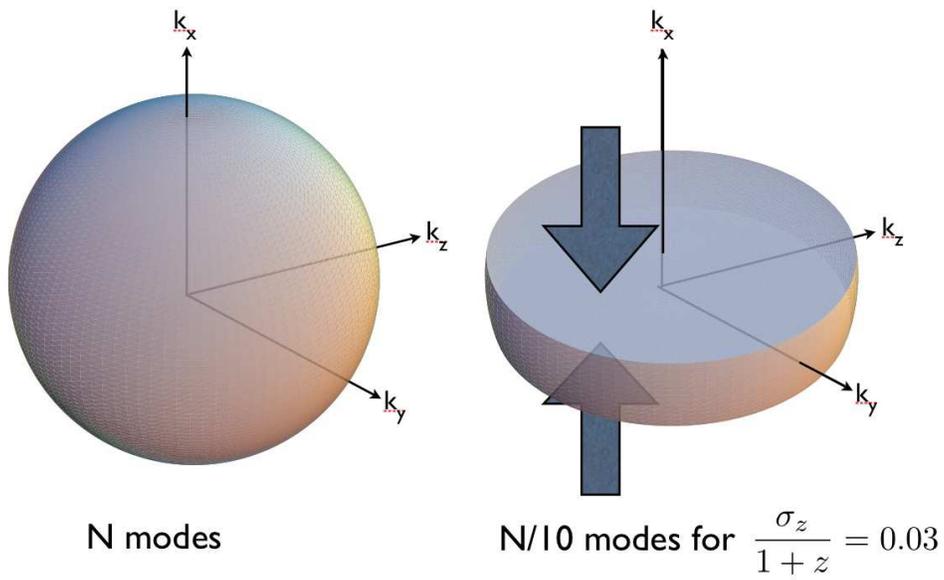,angle=0,width=15cm}
\end{center}
\caption{Illustration of the loss of modes in three dimensional
  Fourier space $(k_x, k_y, k_z)$ by smearing along the $x$ (redshift)
  axis. Only long-wavelength modes $>$ smearing length are
  unsuppressed and the $|k|<0.2$ sphere is truncated to a thin disk.}
\label{figphotexample}
\end{figure*}

Prior to the smearing effect of photometric redshifts, the available
Fourier structure modes in the linear regime comprise a sphere in Fourier
space of radius $k < k_{\rm linear} \approx 0.2 \, h$ Mpc$^{-1}$ (where
the value of $k_{\rm linear}$ depends on redshift). Afterwards, the
damping term $\exp{(-k_x^2 \sigma_x^2)}$ implies that only a thin slice of
this sphere with $|k_x| \lesssim 2/\sigma_x \approx 0.02 \, h$ Mpc$^{-1}$
is able to contribute useful power spectrum signal.  This is illustrated
schematically in Figure~\ref{figphotexample}.  Considering those modes
contributing to a Fourier bin centred about scale $k$, the reduction in
usable Fourier space volume corresponds to a factor $(k \times \sigma_x/2)$. 
At a scale $k = 0.2 \, h$ Mpc$^{-1}$, this represents a loss by a factor
of 10 ($\sigma_0 = 0.03$, $z = 1$).  For the same surveyed area, we can
hence expect the error ranges in the derived power spectrum to worsen by a
factor $\approx \sqrt{10}$ in comparison to a spectroscopic survey
(although also note that the scaling of the error with $k$ also changes
from $\delta P_{\rm spec} \propto k^{-1}$ to $\delta P_{\rm
photo} \propto k^{-1/2}$). The errors on the resulting dark energy parameters will
be worse by a factor of $\simeq \sqrt{20}$ than in a similar sized spectroscopic survey once
one also accounts for the loss on the radial dimension.

Critically, the radial damping due to photometric redshifts results in the
loss of any ability to detect the acoustic oscillations in the {\it
radial} component of the power spectrum (in the above example, the power
spectrum is significantly suppressed for modes with $|k_x| > 2/\sigma_x
\approx 0.02 \, h$ Mpc$^{-1}$, i.e.\ the whole regime containing the
oscillations).  We are only able to apply the standard ruler in the
tangential direction.  As noted earlier, the radial component provides a
direct measure of $H(z)$, contributing significantly to the measurements
of the dark energy model.  Thus in the above example ($\sigma_0 = 0.03$,
$z = 1$) the resulting errors in the dark energy parameters will worsen by
a factor closer to $\approx \sqrt{20}$.  Alternatively, one could
compensate by increasing the survey area (and hence the density of states
in Fourier space) by the corresponding factor.  We note that with
sufficient filter coverage and for special classes of galaxy, the
photometric redshift error $\sigma_0$ may also be reduced to improve dark
energy performance. 

We simulated photo-z surveys using analogous Monte Carlo techniques to
those described in Section \ref{secmeth} (see \cite{BB05} for a more
detailed account).  We introduced a radial Gaussian smearing
\begin{equation}
\sigma_x = \sigma_0 \, (1 + z_{\rm eff}) \, x'(z_{\rm eff})
\end{equation}
into our Poisson-sampled density fields (for a survey slice at redshift
$z_{\rm eff}$).  When measuring the power spectrum we restrict ourselves
to modes with $|k_x| < 2/\sigma_x$ (where the factor of 2 was determined
empirically to be roughly optimal).  The residual damping in the shape of
$P(k)$ is divided out using the known Gaussian damping expression.  We bin
the power spectrum modes in accordance with the total length of the
Fourier vector $k = \sqrt{k_x^2 + k_y^2 + k_z^2}$, noting that only
tangential modes (with $k_x \approx 0$) are being counted.  We fit a 1D
decaying sinusoid (Paper I, equation 3) to the result.  The scatter of the
fitted wavescales across the Monte Carlo realizations is interpreted as
the accuracy of measurement of the quantity $x(z_{\rm eff})/s$ (given that
only tangential modes are involved).  We uniformly populated our survey
volumes such that $n \times P = 3$. 

\begin{figure*}[htbp]
\begin{center}
\epsfig{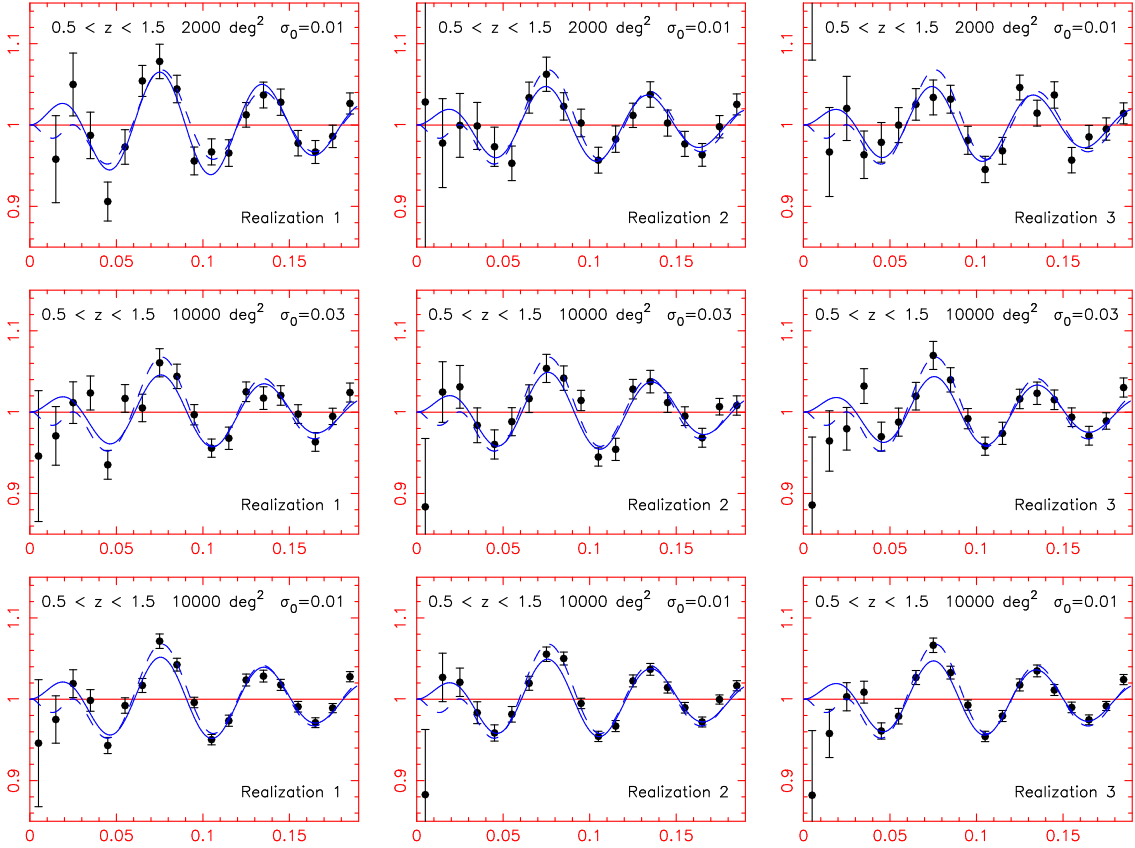}
\end{center}
\caption{Power spectrum realizations for three example photometric
  redshift surveys, varying survey area and photo-z accuracy
  (parameterized by $\sigma_0$).  In all cases we assume a survey
  redshift interval $0.5 < z < 1.5$ (i.e.\ $z_{\rm eff} = 1$).  Only
  radial Fourier modes with $|k_x| < 2/\sigma_x$ (where $\sigma_x =
  \sigma_0 \, (1 + z_{\rm eff}) \, x'(z_{\rm eff})$) are binned when
  measuring $P(k)$, which is plotted divided by a smooth reference
  spectrum.  As in Figure \ref{figpkspec}, the dashed curve is the
  theoretical input $P(k)$ and the solid line is the best fit of our
  simple decaying sinusoidal function (Paper I, equation 3).  The
  $x$-axis is marked in units of $k$ (in $h$ Mpc$^{-1}$) and
  represents the extent of the linear regime at redshift $z_{\rm
    eff}$.  The rows of the Figure represent surveys with parameters
  ($2000$ deg$^2$, $\sigma_0 = 0.01$), ($10{,}000$ deg$^2$, $\sigma_0
  = 0.03$) and ($10{,}000$ deg$^2$, $\sigma_0 = 0.01$); the columns
  display the first three Monte Carlo realizations in each case.}
\label{figpkphot}
\end{figure*}

We consider two different photo-z accuracies: $\sigma_0 = 0.03$,
representing the typical fidelity of the current best photo-z studies, and
$\sigma_0 = 0.01$, which we somewhat arbitrarily adopt as an upper limit
to future improvements.  \cite{BB05} consider a much wider range of
possibilities.  We note that in our methodology, reducing the value of
$\sigma_0$ by some factor is equivalent to covering a proportionately
larger survey area.  Figure \ref{figpkphot} displays some Monte Carlo
realizations of measured power spectra for these photometric redshift
surveys, assuming a redshift range $0.5 < z < 1.5$.  We assume a survey
solid angle of $10{,}000$ deg$^2$, but also consider a smaller project
($2000$ deg$^2$ with $\sigma_0 = 0.01$). 

\begin{figure}[htbp]
\begin{center}
\epsfig{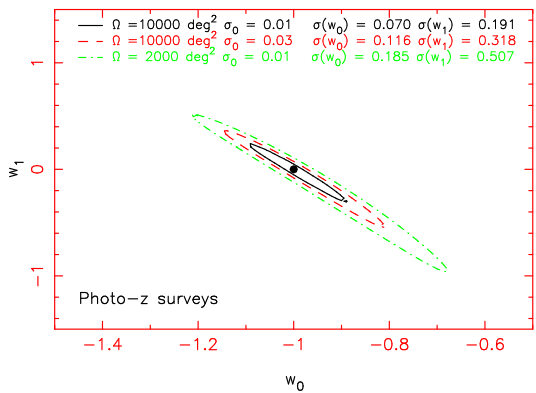}
\end{center}
\caption{Dark energy measurements resulting from three different
  photometric redshift surveys covering the interval $0.5 < z < 3.5$.
  We assume perfect prior knowledge of $(\Omega_{\rm m},h,\Omega_{\rm
    k})$. Table \ref{tabw0w1} lists results for a range of different
  priors.}
\label{figw0w1phot}
\end{figure}

Figure \ref{figw0w1phot} illustrates the resulting measurement of the dark
energy parameters $(w_0,w_1)$ for these survey configurations, assuming
that we can span the redshift range $0.5 < z < 3.5$ (also see Tables
\ref{tabxdx} and \ref{tabw0w1}).  These $(w_0,w_1)$ contours are computed
using the same method as Section \ref{secmethlik}, utilizing measurements
of $x(z)/s$ in three redshift bins of width $\Delta z = 1$ (for real data,
narrower redshift slices would be used and the results co-added).  Each of
these redshift constraints corresponds to a degenerate line in the
$(w_0,w_1)$ plane (i.e.\ constrains one degree of freedom in the dark
energy model) but, as described in Section \ref{secmethlik}, the direction
of degeneracy slowly rotates with redshift: the combination of the
likelihoods for each redshift bin results in closed contours. 

Note that for photometric redshift surveys, the cosmological priors on
$\Omega_{\rm m} h^2$ and $\Omega_{\rm m}$ required to achieve a given
measurement precision of $(w_0,w_1)$ are much tighter than for
spectroscopic surveys (see Table \ref{tabw0w1}).  This is because in the
absence of $H(z)$ information, the photo-$z$ survey must achieve a
significantly tighter measure of $x(z)$ to recover a corresponding
measurement accuracy of the dark energy parameters, rendering it more
susceptible to uncertainties in $\Omega_{\rm m}$ and $h$. 

The real figure-of-merit for comparison of practical instruments is the
accuracy with which the dark energy model can be measured {\it for a fixed
total observing time} or {\it at a fixed cost}.  The myriad details of
comparing large imaging cameras with large spectroscopic systems are
beyond the scope of this paper.  However, we note that the proposed Large
Synoptic Survey Telescope \citep[LSST; ][]{LSST} could image half of the
entire sky to the required depth ($V \approx 25$) in multiple colours
every 25 nights; such a survey would produce dark energy constraints
comparable with a spectroscopic (e.g.\ KAOS) survey of 1000 deg$^2$ (the
latter requiring 170 nights) {\em if $\sigma_0 = 0.01$ could be achieved}. 
We regard this level of photometric redshift accuracy as unlikely for
ground-based surveys. We note that the KAOS measurements additionally
constrain $H(z)$ leading to qualitatively more robust measurement of dark
energy. Further the KAOS survey could be improved by adding more area,
whereas once the LSST has observed the whole celestial hemisphere, there
is obviously no further gain in $w(z)$ information from further passes. 
However, since `all-sky' deep multi-colour surveys are being performed for
other scientific reasons (e.g.\ cosmic shear analysis), it is of
considerable value to utilize these data for baryonic oscillation studies. 
Furthermore, as discussed in detail by \cite{BB05}, a photometric redshift
survey of several thousand square degrees may provide constraints
competitive with spectroscopic surveys in the short term. 

\section{Conclusions}

This study has extended the methodology of Blake \& Glazebrook (2003) to
simulate measurements of the cosmic {\it evolution} of the
equation-of-state of dark energy $w(z)$ from the baryonic oscillations,
using the simple parameterization $w(z) = w_0 + w_1 z$. The methodology
used is very similar to that of Paper I, treating the primordial baryonic
oscillations in the galaxy power spectrum as a standard cosmological
ruler, whilst dividing out the overall shape of the power spectrum in
order to maximize model-independence.  In this study we make the
improvement of fitting independent radial and tangential wavescales,
showing that this is directly equivalent to measuring 
$D_A(z)$ and $H(z)$ in a series of redshift slices in units of the sound horizon.
This results in 
improved constraints upon $(w_0,w_1)$.  We have
tested the approximations encoded in our approach and found them all to be
satisfactory, increasing our confidence in the inferred error
distributions for the dark energy parameters.  The simulated accuracies
for $(w_0,w_1)$ are roughly consistent with other estimates in the
literature based on very different analysis methods. 

Our baseline `KAOS-like' optical surveys of $\sim 1000$ deg$^2$, which can
be realized by the next generation of spectroscopic instruments at
ground-based observatories, deliver measurements of the dark energy
parameters with precision $\Delta w_0 \approx 0.15-0.2$ and $\Delta w_1
\approx 0.3-0.4$.  In statistical terms, these constraints are poorer than
those which may be provided by a future space-based supernova project such
as the SNAP proposal.  We note, however, that the baryonic oscillations
method appears to be substantially free of systematic error, with the
principal limitation being the amount of cosmic volume mapped.  In
addition, any measurement of deviations from a cosmological constant model
is of sufficient importance for physics that {\it entirely independent}
experiments would be demanded to confirm the new model.  A next-generation
radio telescope with a FOV $\approx 100$ deg$^2$ at $1.4$ GHz, performing
a redshift survey of 21cm emission galaxies over several $1000$ deg$^2$,
may be available on a similar timescale. 

We have considered the observational possibilities of more extensive
baryonic oscillation experiments covering a significant fraction of the
whole sky.  Such a survey (encompassing $0.5 < z < 3.5$) may be
straight-forward using a dedicated several-year space mission with
slitless spectroscopy.  In radio wavebands, the Square Kilometre Array
would be able to survey the entire visible sky out to $z \approx 1.5$ in 6
months, if equipped with a sufficiently large FOV.  These experiments
would deliver extremely precise measurements of the dark energy model with
accuracy $\Delta w_0 \approx 0.03-0.05$ and $\Delta w_1 \approx 0.06-0.1$
and would be invaluable to pursue if a significant non-vacuum dark energy
signal was detected by smaller surveys. 

We have also explored in detail the potential of photometric redshift
optical imaging surveys for performing baryonic oscillations experiments. 
The loss of the radial oscillatory signal, due to the damping caused by
the redshift errors, implies that we can no longer recover information
about the Hubble constant at high redshift. However, the baryonic
oscillations can still be measured using tangential Fourier modes.  A deep
$2000$ deg$^2$ imaging survey with excellent photometric redshift
precision ($\sigma_0 < 0.03$) would allow the oscillations to be detected (2.5$\sigma$ significance). 
Useful constraints upon the dark energy model are possible if $\sim
20{,}000$ deg$^2$ can be surveyed (such an experiment with  $\sigma_0 =
0.03$ is roughly
equivalent to a $\sim 1000$ deg$^2$ spectroscopic survey). 

We conclude that the baryonic oscillations in the clustering power
spectrum represent one of the rare accurate probes of the cosmological
model, possessing the potential to delineate cleanly any cosmic evolution
in the equation-of-state of dark energy, via accurate measurements of
$D_A(z)$ and $H(z)$ in a series of redshift slices. Importantly, such an
experiment is likely to be substantially free of systematic error.  Recent
observations of SDSS Luminous Red Galaxies at $z = 0.35$ have provided the
first convincing detection of the acoustic signature and validation of the
technique.  The challenge now is to create the large-scale surveys at
higher redshifts required for mapping the properties of the mysterious
dark energy. 

\acknowledgments

KG and CB acknowledge generous funding from the David and Lucille Packard
foundation and the Center for Astrophysical Sciences, Johns Hopkins
University.  CB warmly thanks his colleagues at the University of New
South Wales where most of this work took place, especially Warrick Couch,
and acknowledges funding from the Australian Research Council.  CB also
thanks Sarah Bridle, Filipe Abdalla and Steve Rawlings for many valuable
conversations.  We are grateful for useful discussions with Dan Eisenstein
and Eric Linder.  CB acknowledges current funding from the Izaak Walton
Killam Memorial Fund for Advanced Studies and the Canadian Institute for
Theoretical Astrophysics. 

\appendix

\section{Tests of approximations}

\subsection{Effect of redshift-space distortions}

As mentioned in Section \ref{secsumm}, a power spectrum measured from a
real redshift survey is liable to suffer systematic shape distortions.  We
performed a test to verify that we were able to divide out such
distortions via a smooth fit, prior to fitting for the acoustic
oscillations.  For a survey covering $0.5 < z < 1.3$ and $1000$ deg$^2$,
we incorporated a redshift-space distortion into each Monte Carlo
realization by smearing the redshift of each simulated galaxy by the
equivalent of a radial velocity of $300$ km s$^{-1}$. This process has the
effect of damping the overall power spectrum by a factor increasing with
$k$ and amounting to about $20\%$ at $k = 0.2 \, h$ Mpc$^{-1}$.  After
division by the `no-wiggles' reference spectrum of \cite{EH98}, we fitted
an additional second-order polynomial to the residual, in order to remove
the extra shape distortion.  We found that the scatter of the fitted
acoustic scales across the Monte Carlo realizations (i.e.\ the precision
with which we could recover the standard ruler) was entirely unchanged by
this more complex fitting procedure. 

\subsection{Streamlined vs Full methodology}

In this Section we evaluate the validity of the approximations employed in
our `streamlined method' (Section \ref{secstream}) with respect to the
full methodology (Section \ref{secsteps}).  These approximations are
implemented because it would otherwise require a prohibitively long
computational time to cover the grid of dark energy parameters $(w_0,w_1)$
with sufficient resolution.  In particular, we wish to test that: 

\begin{enumerate}

\item The constraints upon the dark energy parameters $(w_0,w_1)$
  arising from the fitted radial and tangential power spectrum
  wavescales can be deduced by supposing that these fitted wavescales
  measure the quantities $x/s$ and $x'/s$, respectively.

\item Dark energy measurements arising from a survey of solid angle
  $A_2$ may be inferred from those resulting from a survey of area
  $A_1 < A_2$ by multiplying the accuracy of $x/s$ and $x'/s$ deduced
  from this former survey by a factor $\sqrt{A_1/A_2}$.

\item A reasonably broad survey redshift interval, $z_{\rm min} < z <
  z_{\rm max}$, can be utilized to measure $x/s$ and $x'/s$ at an
  `effective' redshift $z_{\rm eff} = (z_{\rm min} + z_{\rm max})/2$.

\end{enumerate}

In order to test the effect of these approximations, we consider a `test
case' which we analyze using both the streamlined and the full method. 
This test case comprises a spectroscopic survey covering the redshift
interval $0.7 < z < 1.3$ (i.e.\ approximately the KAOS $z \sim 1$
component), but with an increased area of $4000$ deg$^2$. This is the
maximum size we can simulate in a reasonable time. Implementing the
full method, we first divide the redshift range into three bins of width
$\Delta z = 0.2$.  For each bin, we consider a grid of trial dark energy
parameters $(w_0,w_1)$ (spaced by $\Delta w_0 = 0.05$, $\Delta w_1 = 0.1$)
and create 50 Monte Carlo realizations for each grid point, deducing the
relative probability of each trial cosmology by assessing the position of
the theoretical standard ruler scale in the distributions of fitted radial
and tangential wavescales (as described in Section \ref{secsteps}).  We
then obtain the overall constraints upon the dark energy model, resulting
from the full method, by multiplying together these three likelihood maps. 
We compare these measurements with those resulting from the streamlined
approach, based upon scaling a single analysis of a ($0.7 < z < 1.3$,
$1000$ deg$^2$) survey by a factor $\sqrt{4}$, and deeming the effective
redshift to be $z_{\rm eff} = 1$. 

\begin{figure}[htbp]
\begin{center}
\epsfig{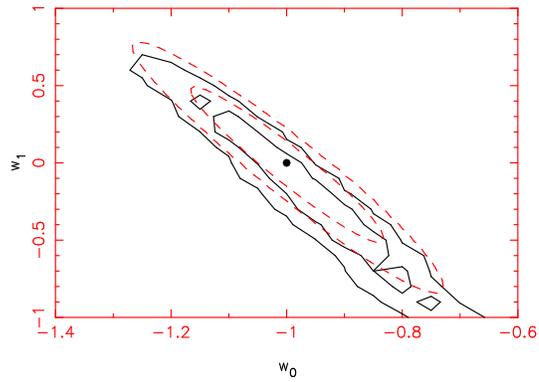}
\end{center}
\caption{Comparison of $(68\%,95\%)$ likelihood contours deduced using
  the `full methodology' (Section \ref{secsteps}; solid lines) and
  `streamlined methodology' (Section \ref{secstream}; dashed lines)
  for a test survey covering area $4000$ deg$^2$ and redshift interval
  $0.7 < z < 1.3$.  The two sets of contours are in reasonable
  agreement.}
\label{figw0w1test}
\end{figure}

These two likelihood maps in the $(w_0,w_1)$ plane are compared in Figure
\ref{figw0w1test}.  The grid is necessarily quite coarse and noisy for the
full methodology, but it is clear that the likelihood patterns are in
reasonable agreement concerning both width and degeneracy direction.  The
most noticeable discrepancy appears in the region $w_0 > -1$, $w_1 < 0$,
where dark energy has a reduced influence at high redshift but the disagreement
is of marginal significance.  In
the regime $w_1 > 0$, the agreement is much better.
We conclude that within the noise of the simulation
the two sets of contours are a good match though it would be desirable in the
future to test this further.

\end{document}